\documentclass[floats,floatfix,amssymb,prd,twocolumn,superscriptaddress,nofootinbib]{revtex4-1}

\usepackage[english]{babel}
\usepackage[utf8]{inputenc}
\usepackage{amsmath}
\usepackage{mathbbol}
\usepackage{amssymb}
\usepackage{bbold}
\usepackage{graphicx,amsfonts}
\usepackage{float}
\usepackage{epsfig}
\usepackage{bm}
\usepackage{mathrsfs}
\usepackage{siunitx}
\usepackage{mathtools}
\usepackage{enumerate}
\usepackage{amsthm}
\usepackage{bbm}
\usepackage{comment}
\usepackage{physics}
\usepackage{url}
\usepackage{upgreek}
\usepackage[svgnames]{xcolor}
\usepackage{multirow}

\usepackage[colorlinks=true,
linkcolor=blue,
urlcolor=blue,
citecolor=blue]{hyperref}
\usepackage{etoolbox}
\makeatletter

\begin{document}
\title{On the universality of late-time ringdown tails}

\author{Romeo Felice Rosato}
\affiliation{Dipartimento di Fisica, Sapienza Università di Roma \& INFN, Sezione di Roma, Piazzale Aldo Moro 5, 00185, Roma, Italy}

\author{Paolo Pani}
\affiliation{Dipartimento di Fisica, Sapienza Università di Roma \& INFN, Sezione di Roma, Piazzale Aldo Moro 5, 00185, Roma, Italy}

\begin{abstract}
    The late-time response of vacuum black holes in General Relativity is notoriously governed by power-law tails arising from the wave scattering off the curved spacetime geometry far from the black hole.
    While it is known that such tails are universal to a certain extent, a precise characterization of their key ingredients is missing.
    Here we provide an analytical proof that the tail fall-off is universal for any effective potential asymptotically decaying as $1/r^2$, 
    while the power-law decay is different if the potential decays as $1/r^\alpha$ with $1<\alpha<2$. This result extends and revises some previous work and is in agreement with numerical analyses.
    Our proof is based on an analytical evaluation of the branch cut contribution to the Green function, and includes charged black holes, different kinds of perturbations, Teukolsky equation for the Kerr metric, exotic compact objects, extensions of General Relativity, and environmental effects. 
    In the latter case, our results indicate that tails are largely insensitive to a wide range of physically motivated matter distributions around black holes, including the Navarro–Frenk–White profile commonly used to model dark matter.
\end{abstract}

\maketitle

\tableofcontents

\section{Introduction}
Binary black hole~(BH) mergers serve as a natural testbed to investigate the full nonlinear dynamics of Einstein’s theory of gravity~\cite{LIGOScientific:2021sio,Berti:2015itd,Berti:2018vdi,Cardoso:2019rvt}. These systems, long studied for their theoretical and astrophysical relevance~\cite{Barack:2018yly}, have gained renewed importance with the emergence of gravitational-wave~(GW) observations~\cite{LIGOScientific:2016aoc}, which now offer direct access to the most extreme regimes of spacetime dynamics.

The evolution of such a binary system typically unfolds in three main phases: a prolonged inspiral, a highly dynamical merger, and a relaxation stage often referred to as the ringdown. During the inspiral, the two BHs orbit each other while gradually losing energy through GW emission and culminating in a rapid coalescence that forms a single, distorted BH. This remnant undergoes a transient phase before settling into a stable, stationary configuration. At intermediate times after the merger, the ringdown is dominated by quasinormal mode oscillations described by a superposition of exponentially damped sinusoids, while at sufficiently late times the signal transitions to a slower, power-law decay known as the GW tail~\cite{Price:1971fb,Price:1972pw}.

These late-time tails arise due to the scattering of GWs off the curved spacetime geometry far from the BH, effectively modifying the propagation of the wave. First discussed in the context of electromagnetic perturbations on a curved background~\cite{DeWitt:1960fc}, they have been predicted for the first time for gravitational perturbations in Price's seminal work on collapsing matter~\cite{Price:1971fb,Price:1972pw}. Such effects are now well understood within the framework of BH perturbation theory, where they are shown to follow specific decay rates depending on the background and the multipole structure of the perturbation~\cite{Cunningham:1978zfa,Cunningham:1979px,Leaver:1986vnb,Gomez:1992,Gundlach:1993tp,Gundlach:1993tn,Ching:1994bd,Burko:1997tb,Barack:1998bw,Bernuzzi:2008rq,Hod:2009my,Poisson:2002jz}.
Detecting this effect is challenging, even in numerical simulations, despite its solid theoretical basis and late-time importance. This difficulty arises because the signal is very faint and requires high precision to distinguish it from errors generated by numerical and boundary artifacts. However, recent studies have shown that incorporating the late-time tail into the ringdown analysis can significantly refine the extraction of quasinormal modes \cite{Thomopoulos:2025nuf}. Furthermore, recent studies suggest that late-time tails may retain imprints of the earlier merger dynamics. In particular, Ref.~\cite{DeAmicis:2024not} presents an analytical investigation of the effects of orbital eccentricity, whose predictions have been largely confirmed by subsequent analyses~\cite{DeAmicis:2024eoy,Ma:2024hzq}. These studies show that, in highly eccentric binaries, the amplitude of the late-time tails can be significantly enhanced, potentially opening the way for the future detection of such effects~\cite{Chiaramello:2020ehz,Albanesi:2023bgi,DeAmicis:2024not}.
Furthermore, a recent study has shown that tails decay more slowly at second order in perturbation theory~\cite{Kehagias:2025xzm,Ling:2025wfv}.

In this work, we focus on these late-time phenomena using a perturbative approach similar to~\cite{Leaver:1986gd,Leaver:1986vnb}, analyzing the Green function of the linearized problem in Fourier space, as done in~\cite{Andersson:1996cm}. We focus on the dominant decay behavior at very late times. In this regime, we demonstrate that the tails follow a universal power-law for a broad class of systems, including
charged (and rotating, using a spherical-harmonics decomposition for the perturbations) BHs, different kinds of perturbations, exotic compact objects, extensions of General Relativity, and environmental effects. 

The independence of the tail decay on the spin of the perturbing field and on the possible presence of electric charge in the BH background was already suggested in~\cite{Gundlach:1993tp, Gundlach:1993tn} in the context of spherical collapse. Corrections to the effective potential have been investigated in the literature, for the first time by~\cite{Ching:1994bd} using the Born approximation to compute corrections on top of a purely centrifugal potential in terms of the tortoise coordinate. 

Here we show that the tail universality is much more easily described in terms of the large-distance behavior of the effective potential in the radial Schwarzschild coordinate. We provide an analytical proof that the tail fall-off is universal for any effective potential asymptotically decaying as $1/r^2$, (where $r$ is the Schwarzschild coordinate), while their power-law decay is different if the potential decays as $1/r^\alpha$ with $1<\alpha<2$.
As we shall discuss, this result extends and revises some previous analyses~\cite{Ching:1994bd}. As we will further demonstrate, the validity of our results applies also to spinning black holes.
The universality of late-time tails has been further investigated in a seminal work \cite{Poisson:2002jz}, which analyzes the propagation of a scalar field on a stationary, asymptotically flat background without imposing any additional symmetries. It is demonstrated that the late-time tail behavior is insensitive to the specific symmetries of the background and is instead entirely determined by the multipolar structure of the perturbation. Our findings are in excellent agreement with those of~\cite{Poisson:2002jz}, despite being obtained through an entirely different formalism.
In the second part of this work, we will discuss the tails in the presence of matter fields around BHs and for compact horizonless objects~\cite{Cardoso:2019rvt}, providing an analytical proof of the tail universality. While it is known that the internal structure of the perturbed object does not affect the tails, to the best of our knowledge this feature has been analyzed only numerically (see, e.g.,~\cite{Bernuzzi:2008rq}). All of our analytical results are supplemented by numerical confirmations obtained solving the corresponding equations in the time domain.
We use the $G=c=1$ geometrized unit system.

\section{Green functions and tails}\label{greenfunctiondiscussion}
\begin{figure}
    \centering
    \includegraphics[width=\linewidth]{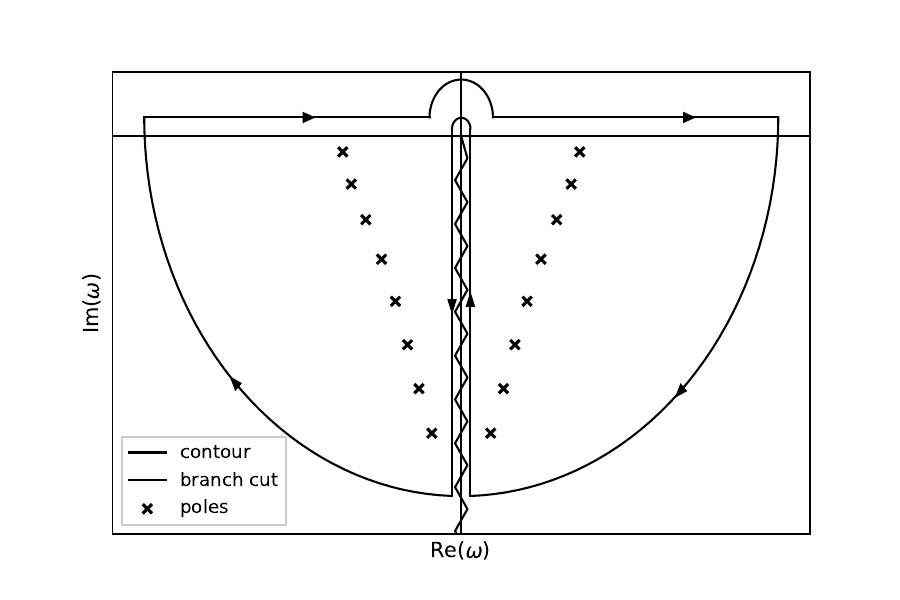}
    \caption{Representation of the contour of integration for Green function in $\omega$ complex plane.}
    \label{fig:contour}
\end{figure}

We will mostly focus on spherically symmetric spacetimes (extension to spinning BHs will be discussed later on).
Within linear perturbation theory, the BH response to an external perturbation in the time domain is described as an inhomogeneous one-dimensional wave equation with a potential
\begin{equation}\label{waveeq}
    \left({\partial^2 \over \partial r_*^2}-{\partial^2 \over \partial t^2}-V_{lm}(r_*)\right)\Psi_{lm}(r_*,t)=S_{lm}(r_*,t)\,
\end{equation}
where $r_*$ is the tortoise coordinate\footnote{For a spherically symmetric metric $ds^2=-h(r)dt^2+dr^2/f(r)+r^2d\Omega^2$, the tortoise coordinate is defined by $dr/dr_*=\sqrt{f(r) h(r)}$. As later discussed, our argument applies to any linear problem that can be modeled by an equation of the form~\eqref{waveeq} regardless of its origin.
}, $(l,m)$ are the spherical-harmonic indices, and $S_{lm}(r_*,t)$ is the source inducing the perturbation. The tails can be studied through the late-time behavior of the Green function $G(r_*,t|r_*',t')$ associated with Eq.~\eqref{waveeq}. If we consider the problem in Fourier space, we can introduce the frequency-domain Green function 
\begin{equation}
    g(r_*,r_*',\omega)=\int_{t'}^\infty dt\,e^{-i\omega(t'-t)}G(r_*,t|r_*',t')\,,
\end{equation}
and the time-domain Green function is recovered by computing the inverse Fourier transform:
\begin{equation}
   G(r_*,t|r_*',t')=\lim_{\epsilon\to0}{1 \over 2\pi}\int_{-\infty+i\epsilon}^{+\infty+i\epsilon} d\omega\,e^{i\omega(t'-t)}g(r_*,r_*',\omega)\,,
\end{equation}
where the shift $i\epsilon$ is needed to avoid the nonpropagating mode at $\omega=0$ in the integration. 
The integral above can be performed in the complex frequency plane, following the contour depicted in Fig~\ref{fig:contour}. 
The frequency-domain Green function has poles in the complex plane (corresponding to the quasinormal modes) and it is a multivalued function, hence a branch cut must be taken into account when computing the integral in the complex frequency plane (see Appendix~\ref{branchcutcomputation}).
Clearly, the Green function in time-domain can be studied as the sum of three different parts: the contribution of the quasinormal frequencies, namely the poles; the boundary at infinity; the integrals around the branch cut frequencies. The latter is the part that contributes to the late-time response~\cite{Leaver:1986gd}. For a BH, this latter contribution reads
\begin{align}\label{eqGreen}
   & G_B^{\rm BH}(r_*,t \mid r'_*,t') = -\frac{1}{2\pi} \int_0^{-i\infty}d\omega\,
e^{-i \omega (t-t')}\psi_{r_-}(r_*<,\omega)\notag\\&
 \Bigg[\left( 
\frac{\psi_{\infty_+}(r_*>,\omega \,e^{2\pi i})}{\mathcal{W}(\omega\,e^{2\pi i})} - \frac{\psi_{\infty_+}(r_*>,\omega)}{\mathcal{W}(\omega)} 
\right) \Bigg]\,,
\end{align}
where $\mathcal{W}=\mathcal{W}(\psi_{r_-},\psi_{\infty_+})$ is the Wronskian of the two solutions $\psi_{r_+}$ and $\psi_{\infty_+}$ of the corresponding homogeneous equation with boundary conditions
\begin{align}
    \psi_{r_-}(r_* \to -\infty)&=e^{-i\omega r_*}\,,\notag\\ 
    \psi_{r_-}(r_* \to +\infty)&=A_{\rm in}(\omega)e^{-i\omega r_*}+A_{\rm out}(\omega)e^{i\omega r_*}\,,
\end{align}
and 
\begin{align}
     \psi_{\infty_+}(r_* \to -\infty)&=B_{\rm in}(\omega)e^{-i\omega r_*}+B_{\rm out}(\omega)e^{i\omega r_*}\,,\notag\\
  \psi_{\infty_+}(r_* \to +\infty)&=e^{i\omega r_*}\,.
\end{align}
Note that Eq.~\eqref{eqGreen} is derived without taking into account any discontinuity of $\psi_{r_-}$ across the branch cut. In Appendix~\ref{branchcutapp}, we show why this is justified. 

We notice that~\cite{Leaver:1986gd}
\begin{equation}\label{psirminus}
    \psi_{r_-}=A_{\rm out}(\omega) \psi_{\infty_+}+A_{\rm in}(\omega) \psi_{\infty_-}\,,
\end{equation}
where $\psi_{\infty_-}=\left(\psi_{\infty_+}\right)^*$, satisfying purely ingoing boundary conditions at infinity. Introducing $K(\omega)$ to satisfy 
\begin{equation}\label{branchcutpsiplus}
   \psi_{\infty_+}(r_*, \omega e^{2\pi i}) = \psi_{\infty_+}(r_*, \omega) - K(\omega) \psi_{\infty-}(r_*, \omega)\,,
\end{equation}
Eq.~\eqref{eqGreen} reads
\begin{align}\label{greenAinAout}
    &G_B^{\rm BH}(r_*, t \mid r'_*, t') = \frac{1}{2\pi } \int_0^{-i\infty} d\omega\,e^{-i\omega(t - t')} K(\omega)  \notag\\ &\quad\quad\Bigg[\frac{\psi_{r_-}(r_*<, \omega) \psi_{r_-}(r_*> , \omega)}{2i\omega A_{\text{in}}(\omega) \left[A_{\text{in}}(\omega) + K(\omega) A_{\text{out}}(\omega)\right]}\Bigg] \,.
\end{align}

\subsection{Small-frequency expansion}
Since we are interested in the late-time decay tails, we focus on the limit $t - t' \gg r_*, r_*'$ and $t - t' \gg 2M$. In this regime, in Eq.~\eqref{eqGreen}, the factor $e^{-i\omega(t-t')}$ is strongly suppressed for any $\omega \neq 0$, and thus the integral is dominated by the behavior near $\omega \to 0$. 
In the $\omega \to 0$ limit, our equations considerably simplify. Rewriting the homogeneous part of Eq.~\eqref{waveeq} in terms of the $r$ coordinate, for a vacuum spherically symmetric background\footnote{In fact, our argument also applies to non-vacuum spherically symmetric backgrounds of the form
\begin{equation}\label{matter_background}
    ds^2 = -h(r)dt^2 + \frac{dr^2}{f(r)} + r^2 d\Omega^2\,,
\end{equation}
however, for simplicity, we temporarily focus on the background in Eq.~\eqref{vacuum_background}. The generalization to the background in Eq.~\eqref{matter_background} will be discussed in Sec.~\ref{generalization}.}
\begin{equation}\label{vacuum_background}
    ds^2 = -f(r)dt^2 + \frac{dr^2}{f(r)} + r^2 d\Omega^2\,,
\end{equation}
we get
\begin{align}\label{waveq_r}
    &\Big(f^2 \frac{d^2}{dr^2} + ff'\frac{d}{dr}  + \omega^2 - V_{lm}(r) \Big)\Psi_{lm} = 0\,.
\end{align}

In order to account for a wide range of physical cases, we consider a parametrized effective potential of the form
\begin{equation}\label{mod_pot}
    V_{lm}(r) = f(r) \left( \frac{l(l+1)}{r^2} + \frac{A M^{\alpha-2}}{r^\alpha} + \frac{B M^{\beta-2}}{r^\beta} \right)\,,
\end{equation}
where $A$, $\alpha$, $B$, and $\beta$ are dimensionless constants. We introduce two distinct additive terms to directly match some interesting physical cases, such as
\begin{itemize}
    \item {\it Schwarzschild geometry}: with $f(r) = 1 - 2M/r$, $A = 2(1-s^2)$, $B = 0$, and $\alpha = 3$ for a spin-$s$ perturbation (for $s=2$ this includes only the odd-parity case, see below for the even-parity case).
    \item {\it Reissner--Nordström geometry}: with $f(r) = 1 - 2M/r + Q^2/r^2$, $A = 2(1-s^2)$, $B = -2Q^2(1-s^2)/M^2$, $\alpha = 3$, and $\beta = 4$, again for a spin-$s$ perturbation (odd-parity sector for $s=2$) and where $Q$ is the BH charge.
\end{itemize}
In general, we require that Eq.~\eqref{waveq_r} admits plane waves as asymptotic solutions, which implies\footnote{This can happen only if the potential satisfies $rV_{lm}(r)\to0$ when $r\to\infty$, which excludes $\alpha,\beta=1$. Therefore, we will assume from now on that $\alpha,\beta>1$.}
\begin{equation}
    \alpha>1 \quad \quad \quad \beta>1 \quad \quad \quad  \lim_{r\to\infty}{ f(r) \over r} =0 \,.
\end{equation} 

It is useful to manipulate Eq.~\eqref{waveq_r} by introducing the redefinition
\begin{equation}\label{Psivsy}
\Psi_{lm} = \left(f(r)\right)^{-2iM\omega} y_{lm}(r)\,,
\end{equation}
and changing variables to $z = \omega r$. In the $\omega \to 0$ limit and keeping $z$ finite, Eq.~\eqref{waveq_r} becomes
\begin{align}
     &\left[f(z) \frac{d^2}{dz^2} + f'(z) \frac{d}{dz} -  \left(A\, \frac{\omega^{\alpha-2}}{z^{\alpha+1}} + B\, \frac{\omega^{\beta-2}}{z^{\beta+1}} \right.\right.\notag\\&\quad \quad \quad\left.\left.+ \frac{l(l+1)}{z^2} - \frac{1}{f(z)}\right)\right] y_{lm}(z) + \mathcal{O}(\omega) = 0\,.
\end{align}

In asymptotically flat spacetimes, where $f(r) \to 1$ as $r \to \infty$, we can expand
\begin{equation}
    f(r) = 1 + \sum_{n>0} c_n r^{-n}\,,
\end{equation}
with $c_1 = -2M$. In terms of the $z$ variable, this becomes
\begin{equation}
    f(z) = 1 + \sum_{n>0} c_n \omega^n z^{-n}\,.
\end{equation}
Hence,
\begin{equation}
    \lim_{ M\omega \to 0} f(z) = 1\,, \quad\quad\quad \lim_{M\omega \to 0} f'(z) = 0\,,
\end{equation}
as long as $z$ is kept finite. This implies that the equation we need to solve simplifies to
\begin{align}\label{eqZpsi}
     &\left[{d^2 \over dz^2}-\Big(A\, {(M\omega)^{\alpha-2}\over z^{\alpha}}+B\, {(M\omega)^{\beta-2}\over z^{\beta}}+\, \right.\notag\\&\left.\quad\quad\quad\quad\quad\quad\quad\quad\,+\frac{l (l+1)}{z^2}-1\Big)\right]y_{lm}(z)=0\,.
\end{align}
Provided $\alpha,\beta> 2$ the equations simply reads
\begin{equation}
    \left[{d^2 \over dz^2}+1-\frac{l (l+1)}{z^2}\right]y_{lm}(z)=0\,,
\end{equation}
The cases $1<\alpha,\beta\leq 2$ are discussed in Sec.~\ref{bhcase}. Since the above equation does not depend on $A$ and $B$, it is already evident that its solutions will be universal. This equation is a special case of a Coulomb wave equation and admits as a pair of independent solutions
\begin{equation}\label{setofindependentsolutions}
    y_{lm,1}=F_{l}(0,z) \quad\quad\quad y_{lm,2}=G_{l}(0,z)\,,
\end{equation}
which can be connected to Bessel functions $J_\alpha,Y_\alpha$ as
\begin{align}
    y_{lm,1}=\left({\pi z \over2}\right)^{1 \over 2} J_{l+{1 \over 2}}(z) \,,\quad y_{lm,2}=-\left({\pi z \over2}\right)^{1 \over 2} Y_{l+{1 \over 2}}(z)\,.
\end{align}
Given this set of independent solutions, we now need to  reconstruct $\psi_{r+}$ and $\psi_{\infty_+}$ appearing in Eq.~\eqref{eqGreen} and satisfying the appropriate boundary conditions at $r_*\to\pm\infty$.

The limit $r_* \to -\infty$ corresponds to approaching a finite value of $r = r_h$, defined by the condition $f(r_h) = 0$. Since $r_h$ is finite, the corresponding value $z_h = \omega r_h$ tends to zero as $\omega \to 0$. In this limit, the solutions appearing in Eq.~\eqref{setofindependentsolutions} take the form
\begin{equation}
     y_{lm,1} \to {1 \over \left(2l+1 \right)!!}z^{l+1} \qquad   y_{lm,2} \to {\left(2l+1 \right)!! \over 2l+1}z^{-l}\,.
\end{equation}
Given the definition introduced in Eq.~\eqref{Psivsy}, the incoming wave behavior at $r_* \to -\infty$ has already been factored out. As a result, among the set of independent solutions, only $y_{lm,1}$ correctly describes a purely ingoing wave in the $\omega \to 0$ limit, since $y_{lm,1}$ remains regular as $z \to 0$, whereas $y_{lm,2}$ diverges in this limit. With an appropriate normalization, we thus find, in terms of the original $r$ coordinate,
\begin{align}\label{psirminus_F}
&\lim_{M\omega \to 0} \psi_{r_-} (r,\omega) \sim \notag\\&\quad\quad\sim (2l+1)!!(\omega)^{-l-1} \left( f(r)\right)^{-2iM\omega} F_{l}(0,r\omega)\,.
\end{align}

In the limit $r_* \to \infty$, $r \sim r_*$ since $f(r) \to 1$ asymptotically. Coulomb wave functions exhibit the following limiting behavior as $z \to \infty$
\begin{equation}
    F_{l}(0,z) \pm i G_{l}(0,z)=H^\pm_{l}(0,z) \sim e^{\pm i z} e^{\mp i{\pi l \over 2}}\,.
\end{equation}
Hence we can set 
\begin{align}
    &\lim_{M\omega \to 0} \psi_{\infty \pm} (r,\omega) \sim \notag\\ & \quad\sim\left(f(r)\right)^{-2iM\omega} 
\left[ G_{l}(0, r\omega) \pm i F_{\nu}(0, r\omega) \right] e^{ \pm i {\pi l \over 2}}\,,
\end{align}
in order to match the correct asymptotic behaviors.

\subsection{Branch cut} 
In Eq.~\eqref{branchcutpsiplus}, we introduced the function $K(\omega)$. Considering the solutions derived above at the leading order in $M\omega$, that is, neglecting $\mathcal{O}(M\omega)$ terms in the equation, one finds that $K(\omega) =0$. Therefore, to correctly capture the discontinuity across the branch cut, it is necessary to retain subleading corrections in $M\omega$.\\
The derivation of this result is nontrivial and, for the sake of clarity and conciseness, it has been deferred to the appendix. In Appendix~\ref{branchcutcomputation}, we show that for all $\alpha, \beta > 2$, the function $K(\omega)$ is given by
\begin{equation}\label{kvalue}
    K(\omega) = -4\pi M \omega (-1)^l\,,
\end{equation}
which is again independent of $A$ and $B$.

\subsection{Green function for $\omega\to0$} 
The computation of $A_{\rm in}(\omega)$ and $A_{\rm out}(\omega)$ entering Eq.~\eqref{greenAinAout} requires the solution of the following system of equations (that can be solved for any value of $r$)
\begin{equation}
    \begin{cases}
        \psi_{r_-}=A_{\rm in}\psi_{\infty-}+A_{\rm out}\psi_{\infty_+}\\
        \partial_r \psi_{r_-}=A_{\rm in}\partial_r \psi_{\infty-}+A_{\rm out}\partial_r \psi_{\infty_+}
    \end{cases}\,.
\end{equation}
 Proceeding in this way, one obtains 
\begin{align}
 \lim_{\omega \to 0} A_{\text{out}}(\omega) \sim - (-1)^l\lim_{\omega \to 0} A_{\text{in}}(\omega) \\ \lim_{\omega \to 0} 2i\omega A_{\text{in}}(\omega) \sim -(2l+1)!!(i)^l \omega^{-l}\,.   
\end{align}

Supposing $r_* > r_*'$ (corresponding to the observer being farther from the source region), in the $\omega \to 0$ limit Eq.~\eqref{greenAinAout} reduces to
\begin{align}\label{Greenomegalow}
    G_B^{\rm BH}&(r_*, t  \mid r'_*, t')\sim 2iM \int_0^{-i\infty} d\omega\,\Big(f(r)\notag\\&f(r')\Big)^{-2iM\omega}  F_l(0, \omega r) F_l(0, \omega r') e^{-i\omega(t - t')} \,.
\end{align}
We can introduce the retarded time
\begin{equation}\label{retardedtime}
t_r=t+2M\log(f(r))\,,\,\,t'_r=t'-2M\log(f(r'))\,,
\end{equation}
so that Eq.~\eqref{Greenomegalow} reads
\begin{align}\label{Greenomegalow_ret}
    G_B^{\rm BH}&(r_*, t  \mid r'_*, t')\sim \notag\\\sim&2iM \int_0^{-i\infty}F_l(0, \omega r) F_l(0, \omega r') e^{-i\omega(t_r - t_r')} d\omega \,.
\end{align}

We stress that if $r$ and $r'$ are not too close to the horizon radius $r_h$, that is, the source not being localized at the horizon and the observer sufficiently far from it, then $t_r\simeq t$ and $t_r'\simeq t'$.
We can now focus on two interesting limits of the previous expression~\cite{Leaver:1986gd}.

Firstly, we consider the regime where $r_* - r_*' \ll t - t'$ and $t - t' - r_* + r_*' \ll r_*$, often called {\it future null infinity} ($\mathscr{I}_+$). These conditions are equivalent to requiring that $t - r_* \gg t' - r_*'$ and $t \gg t' - r_*'$, meaning that the source is localized within finite times and distances, while the observer is located at a much larger distance. 
In this limit, we have $z' \to 0$ and $z \to \infty$, so that
\begin{equation}\label{asymptoticform}
    F_l(0, \omega r') \sim \frac{(\omega r')^{l+1}}{(2l+1)!!}\,,\, F_l(0, \omega r) \sim \sin\left(\omega r - \frac{l\pi}{2}\right),
\end{equation}
where we used the behavior of the Coulomb wavefunctions. 
Integrating by parts $l+1$ times, one then obtains\footnote{Notice that $\sin\left(\omega r - \frac{l\pi}{2}\right) = \frac{1}{2i}\left(e^{i\omega r - il\pi/2} - e^{-i\omega r + il\pi/2}\right)$, but only the first exponential contributes, as the second one leads to a much faster decaying tail.}
\begin{equation}
    G_{B,\mathscr{I}_+}^{\rm BH}(r_*, t \mid r'_*, t') 
\sim M A_{l} {(r')^{l+1}\over  (t_r - t'_r - r_*)^{l+2}}\,\,.
\end{equation}
with $A_l=(-1)^{l+1} \frac{2(l+1)!}{(2l+1)!!}$. This perfectly coincides with Leaver's result~\cite{Leaver:1986gd} obtained in the context of a Schwarzschild BH. However, here it has been derived in a much more general framework, with few requirements. 

Secondly, we consider the regime where $r_* \ll t - t'$ and $r_*' \ll t - t'$, often called {\it timelike infinity} ($i_+$). This case leads to the very-late-time, nonradiating tails, first discovered by Price~\cite{Price:1971fb,Price:1972pw} for extended source fields. In this limit, we have
\begin{equation}\label{asymptoticformBis}
    F_l(0, \omega r') \sim \frac{(\omega r')^{l+1}}{(2l+1)!!}\,, \quad F_l(0, \omega r) \sim \frac{(\omega r)^{l+1}}{(2l+1)!!}\,,
\end{equation}
which yields
\begin{align}
G_{B,\mathscr{I}_+}(r_*, t \mid r'_*, t')\sim  M B_l {(rr')^{l+1} \over  (t_r - t'_r)^{2l+3}}\,.
\end{align}
with $B_l=(-1)^{l+1} \frac{2(2l+2)!}{\left[(2l+1)!!\right]^2}$. Clearly, for $t \gg 2M$, this implies a power-law tail of the form $\propto t^{-2l - 3}$. Therefore, we have shown that the tail is universal for any potential in the form~\eqref{mod_pot} with $\alpha,\beta>2$. Below we generalize this proof.
It is important to emphasize that our derivation is based exclusively on the $M\omega \to 0$ expansion, without assuming a large-distance expansion, differently from previous studies.

\subsection{Generalization of the proof}\label{generalization}
We emphasize that the procedure we followed actually applies to a class of equations broader than the form given in Eq.~\eqref{waveq_r}. In particular, we may consider the more general equation:
\begin{align}\label{waveq_rbis}
    \left(a(r)\frac{d^2}{dr^2} + b(r)\frac{d}{dr} + \omega^2 - V_{lm}(r)\right) \Psi_{lm} = 0\,.
\end{align}
Upon changing variables to \( z = \omega r \), the leading-order behavior in the \( M\omega \to 0 \) limit remains governed by Eq.~\eqref{eqZpsi}, provided that the following conditions are satisfied:
\begin{equation} \label{conditions}
    \lim_{\omega \to 0} a(z) = 1 \quad \quad \lim_{\omega \to 0} \frac{b(z)}{\omega} = 0\,.
\end{equation}
Under these mild assumptions, the entire subsequent analysis continues to hold. Eq.~\eqref{waveq_rbis} clearly encompasses the case of a spherically symmetric vacuum solution $ds^2=-f(r)dt^2+dr^2/f(r)+r^2 d\Omega^2$, by choosing $a(r)=f^2(r)$ and $b(r)=f(r)f'(r)$. However, non-vacuum spherically symmetric solutions (or BH solutions beyond General Relativity) of the form 
\begin{equation}
    ds^2=-h(r)dt^2+{1 \over f(r)}dr^2+r^2 d\Omega^2
\end{equation} are also included. In this case, for many types of perturbations~\cite{Pani:2018inf}
\begin{equation}
    a(r)=f(r)h(r)\,,\,\,b(r)={1 \over 2}\left(f'(r)h(r)+f(r)h'(r)\right)\,.
\end{equation}
As long as $f(r)=1+\mathcal{O}(M/r)$ and $h(r)=1+\mathcal{O}(M/r)$ asymptotically, our discussion remains valid, since the conditions in Eq.~\eqref{conditions} are satisfied.

Besides possible modifications to the structure of the metric, we may also observe that any potential that decays \textit{asymptotically} faster than $1/r^2$ can be considered. Indeed, one may generalize the analysis to a potential of the form
\begin{equation}
    V_{lm}(r) = f(r) \left( \frac{l(l+1)}{r^2} + g(r) \right)\,,
\end{equation}
where $g(r)$ encodes additional subleading terms. Upon changing variables to $z = \omega r$, we are interested in finite values of $z$ when computing the tail behavior. If
\begin{equation}\label{cond}
    \lim_{\omega \to 0} \omega^2 g\left( \frac{z}{\omega} \right) = 0\,,
\end{equation}
then all the steps of our derivation still apply. This condition is clearly satisfied if $r^2 g(r) \to 0$ as $r \to \infty$. For instance, this is the case whenever $g(r) \sim \text{const}/r^\alpha$ for $r \to \infty$ with $\alpha > 2$. More generally, the result holds for any function $g(r)$ satisfying the condition in Eq.~\eqref{cond}. 
This also covers polar-parity perturbations of Reissner-Nordstr\"om BHs, which the previous argument did not include.

\subsection{Comparison with previous work}
 Corrections to the centrifugal potential in the context of late-time tails have been studied in the past~\cite{Ching:1994bd,Hod:2000ya,Hod:2009my}. A detailed analysis was performed in~\cite{Ching:1994bd} using the Born approximation to treat perturbations of a purely centrifugal potential. The goal of this section is to highlight differences between previous work and the present analysis. In~\cite{Ching:1994bd} the problem is analyzed in the tortoise coordinate $r_*$, by considering the following asymptotic form for the potential
 \begin{equation}
     V(r_*)={l(l+1) \over r_*^2}+j_{\alpha}(r_*)\,, \label{dVChing}
 \end{equation}
where $j_{\alpha}(r_*)$ represents a (small) modification with respect to the pure centrifugal potential and the previous expression is considered for large $r_*$. 
In particular, in Ref.~\cite{Ching:1994bd} the effect of the correction term \(j_\alpha(r_*)\) is computed using a Born approximation at large $r_*$, treating it as a perturbation over the exact solutions of the equation with the centrifugal potential alone.
The classes of corrections considered in~\cite{Ching:1994bd} are listed in Table~\ref{tab:Ching}.

\begin{table}[h]
    \centering
    \renewcommand{\arraystretch}{2.0}
    \setlength{\tabcolsep}{4pt}
    \begin{tabular}{|c|c|c|}
        \hline
        \parbox{3cm}{\centering \vspace{4pt}
        Potential\\ correction $j_\alpha(r)$\\ (see Eq.~\eqref{dVChing})
        \vspace{4pt}}
        &
        \parbox{1.7cm}{\centering $\alpha > 2$}
        &
        \parbox{3cm}{\centering \vspace{4pt}
        Late-time\\ behavior (Ref.~\cite{Ching:1994bd})
        \vspace{4pt}}
        \\
        \hline
        \multirow{3}{*}{$\displaystyle \frac{r_*^{\alpha-2}}{r_*^\alpha}$}
        &
        \parbox{1.7cm}{\centering \vspace{4pt}
        odd integer\\ $< 2l+3$
        \vspace{4pt}}
        &
        $\displaystyle t^{-\mu},\quad \mu > 2l+\alpha$
        \\[6pt]
        \cline{2-3}
        &
        \parbox{1.7cm}{\centering \vspace{4pt}
        all other real\\ $\alpha$
        \vspace{4pt}}
        &
        $\displaystyle t^{-(2l+\alpha)}$
        \\[6pt]
        \hline
        \multirow{3}{*}{$\displaystyle \frac{r_*^{\alpha-2}}{r_*^\alpha}
        \log\left( \frac{r_*}{r_{*0}} \right)$}
        &
        \parbox{1.7cm}{\centering \vspace{4pt}
        odd integer\\ $< 2l+3$
        \vspace{4pt}}
        &
        $\displaystyle t^{-(2l+\alpha)}$
        \\[6pt]
        \cline{2-3}
        &
        \parbox{1.7cm}{\centering \vspace{4pt}
        all other real\\ $\alpha$
        \vspace{4pt}}
        &
        $\displaystyle t^{-(2l+\alpha)} \log t$
        \\[6pt]
        \hline
    \end{tabular}
    \caption{Late-time behavior as found in Ref.~\cite{Ching:1994bd}, based on corrections to the potential in $r_*$.}
    \label{tab:Ching}
\end{table}

On the other hand, in our analysis the potential is modified directly in the coordinate $r$ and including an overlall redshift factor, 
\begin{equation}
    V(r) = f(r) \left( \frac{l(l+1)}{r^2} + g_{\alpha}(r) \right)\,. \label{dVour}
\end{equation}
The corresponding late-time behaviors of the perturbations as found in the previous sections are summarized in Table~\ref{tab:OurResult}.

\begin{table}[h]
    \centering
    \renewcommand{\arraystretch}{2.0}
    \setlength{\tabcolsep}{8pt}
    \begin{tabular}{|c|c|}
        \hline
        \parbox{3.2cm}{\centering \vspace{4pt}
        Potential\\ correction $g_\alpha(r)$\\ (see Eq.~\eqref{dVour})
        \vspace{4pt}}
        &
        \parbox{3.8cm}{\centering \vspace{4pt}
        Late-time\\ behavior (this work)
        \vspace{4pt}}
        \\
        \hline
        $\displaystyle \frac{M^{\alpha-2}}{r^\alpha}$
        &
        $\displaystyle t^{-(2l+3)}$
        \\[6pt]
        \hline
        $\displaystyle \frac{M^{\alpha-2}\log(r/2M)}{r^\alpha}$
        &
        $\displaystyle t^{-(2l+3)}$
        \\[6pt]
        \hline
    \end{tabular}
    \caption{Late-time behavior according to the analysis presented in this work, for modifications to the potential in $r$.}
    \label{tab:OurResult}
\end{table}

A comparison of the two results would reveal that for potentials with corrections decaying faster than $1/r_*^3$, specifically for $\alpha > 3$, our results are in full agreement with those of Ref.~\cite{Ching:1994bd}. However, since the transformation from the radial coordinate $r$ to the tortoise coordinate $r_*$ introduces nontrivial corrections, a careful analysis is required to properly compare the two results. As an illustrative example, consider the Schwarzschild geometry, where
\begin{equation}
    \lim_{r_* \to \infty}r(r_*) \to r_*-2M\log\left({r_* \over 2M}\right)+\mathcal{O}(1)\,.
\end{equation}\\
This implies that the centrifugal barrier contains logarithmic terms when expressed in the tortoise coordinate
\begin{align}\label{centrbarrier}
   & \lim_{r_*\to\infty} {l(l+1) \over r^2} \to {l(l+1) \over r_*^2}\notag\\& \quad \quad +{4Ml(l+1) \log\left({r_* \over 2M}\right)\over r_*^3}+ \mathcal{O}\left({\log\left({r_* \over 2M}\right)^2\over r_*^4}\right)\,.
\end{align}
Hence, as shown in Ref.~\cite{Ching:1994bd} and summarized in Table~\ref{tab:Ching}, a pure centrifugal barrier results in a \(t^{-2l-3}\) decay. Furthermore, in the presence of a correction \(g_{\alpha}(r)\) with \(\alpha \geq 3\), it holds that \footnote{If \(\alpha = 3\) and \(\beta = 1\), a multiplicative prefactor will clearly appear in front of the term \(\log(r_*/2M)/r_*^3\). However, this does not affect the overall argument.}
\begin{align}\label{centr+corrections}
  & \lim_{r_*\to\infty} {l(l+1) \over r^2} +A{M^{2-\alpha}\log^\beta\left({r \over 2M}\right) \over r^{\alpha}} \to {l(l+1) \over r_*^2}\notag\\&\quad +{4Ml(l+1) \log\left({r_* \over 2M}\right)\over r_*^3}+ \mathcal{O}\left({\log^\beta\left({r_* \over 2M}\right) \over r_*^{\alpha}}\right)\,,
\end{align}
where, for the sake of generality, we allow \(\beta = 0,1\) to respectively exclude or include a logarithmic term in \(g_{\alpha}(r)\). As shown in Eq.~\eqref{centr+corrections}, the tortoise coordinate expansion introduces only subleading terms, suppressed with respect to \(\log(r_*/2M)/r_*^3\). Consequently, both Ref.~\cite{Ching:1994bd} and our analysis consistently predict a \(t^{-2l-3}\) decay in this regime.

However, for potentials with \(2 < \alpha < 3\), discrepancies arise between our predictions and those of Ref.~\cite{Ching:1994bd}. In particular, while their analysis suggests a modified tail behavior, our approach yields a robust \(t^{-2l-3}\) falloff for any \(\alpha > 2\). Indeed, when \(2 < \alpha < 3\), we find

\begin{align}
  & \lim_{r_*\to\infty} {l(l+1) \over r^2} +A{M^{2-\alpha}\log^\beta\left({r \over 2M}\right) \over r^{\alpha}} \to {l(l+1) \over r_*^2}\notag\\&\quad \quad \quad +A{M^{2-\alpha}\log^\beta\left({r_* \over 2M}\right) \over r_*^{\alpha}}+ \mathcal{O}\left({\log\left({r_* \over 2M}\right)\over r_*^3}\right)\,,
\end{align}
where again, for the sake of generality, we are considering $\beta=0,1$ to exclude/include a subleading logarithmic term. Following Ref.~\cite{Ching:1994bd}, the tail should be modified in this case (see Table~\ref{tab:Ching}), however our analysis suggests that this does not happen. 
\begin{figure}
    \centering \includegraphics[width=1.02\linewidth]{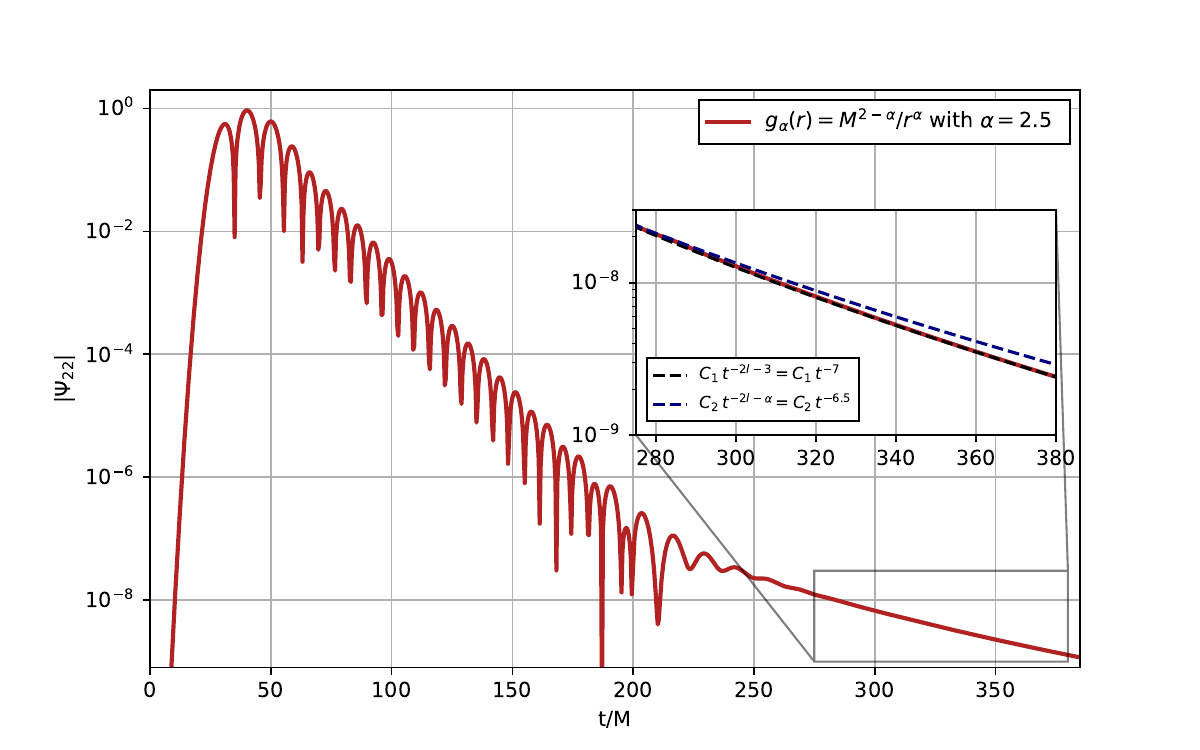} \includegraphics[width=1.02\linewidth]{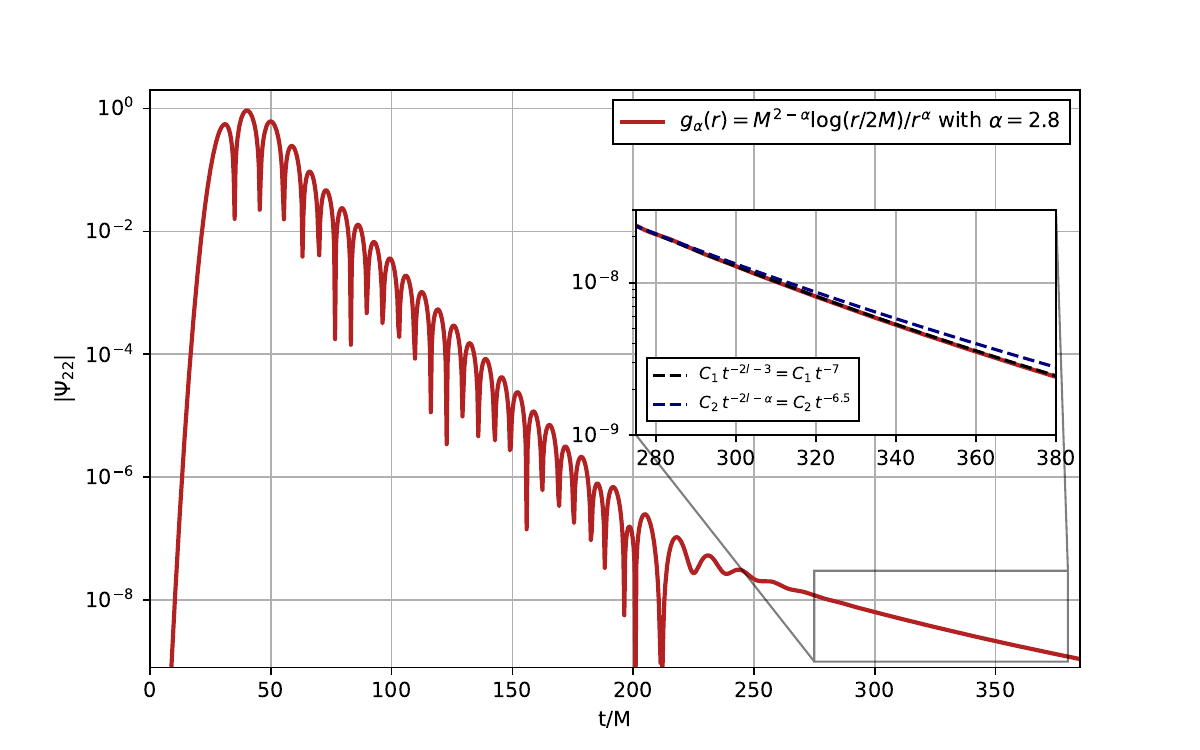}
    \caption{
    Evolution of a Gaussian packet for different configurations of Eq.~\eqref{waveeq} (see Appendix~\ref{Numerics}). The (2,2) waveform multipole is showed for two different corrections to the pure centrifugal potential: i)~$g_\alpha=M^{\alpha-2}/r^{\alpha}$ with $\alpha=2.5$ (upper panel); (ii)~a correction $g_\alpha=M^{\alpha-2}\log(r/(2M))/r^{\alpha}$ with $\alpha=2.8$ (lower panel). In both cases the predictions of Ref.~\cite{Ching:1994bd} do not coincide with our result, as can be better appreciated by the comparison in the insets. Our prediction correctly describes the exact numerical behavior.}
    \label{fig:alpha23}
\end{figure}
We show two examples in Fig.~\ref{fig:alpha23}, where we have verified this behavior through explicit numerical computations in the time domain (see Appendix~\ref{Numerics}), confirming the validity of our result within the framework and approximations adopted in this work. 

\subsection{Extension to spinning BHs}\label{spinningcase}
In the case of a Kerr BH, the analysis must be carried out with care. In Boyer-Lindquist coordinates, the line element for the Kerr spacetime  takes the form
\begin{align}
&ds^2 = - \left(1 - \frac{2Mr}{\Sigma} \right) dt^2 + \frac{\Sigma}{\Delta} dr^2 
- \frac{4Mr}{\Sigma} a \sin^2\theta\, dt\, d\phi \notag \\
&+ \Sigma\, d\theta^2 + \left[ (r^2 + a^2)\sin^2\theta + \frac{2Mr}{\Sigma} a^2 \sin^4\theta \right] d\phi^2\,, \label{eq:kerr_metric}
\end{align}
where \( \Sigma = r^2 + a^2 \cos^2\theta \), \( \Delta = r^2 + a^2 - 2Mr \), whereas \( M \) and \( J = aM \) denote the BH mass and angular momentum. The horizon is located at $r_+=M+\sqrt{M^2-a^2}$. Scalar, electromagnetic, and gravitational perturbations in the exterior region of the Kerr spacetime are described by the Teukolsky master equations~\cite{Teukolsky:1972my,Press:1973zz,Teukolsky:1974yv}
\begin{align}\label{eq:teuk_radial}
&\Delta^{-s} \frac{d}{dr} \left( \Delta^{s+1} \frac{d\, {}_s R_{lm}}{dr} \right)+
\notag\\&\quad \Bigg[ \frac{ K^2 -2is(r - M)K}{\Delta} + 4is\omega r - \lambda
\Bigg] {}_s R_{lm} = 0 \,, 
\end{align}
\begin{align}\label{eq:teuk_angular}
&\left[ (1 - x^2) \frac{d}{dx} \left( \, {}_s S_{lm,x} \right) \right]+  \Big[ (a\omega x)^2 - 2a\omega s x+\notag\\&\quad\quad\quad\quad\quad\quad\quad
  s
+ {}_s A_{lm} - \frac{(m + sx)^2}{1 - x^2} \Big] {}_s S_{lm} = 0 \,, 
\end{align}
where ${}_s S_{lm}(\theta) e^{im\phi}$ are the spin-weighted spheroidal harmonics, $x = \cos\theta$ and $K = (r^2 + a^2)\omega - am$. Moreover, the separation constant $\lambda$ is related to the eigeinvalues ${}_s A_{lm}$ of the angular equation by \(\lambda = {}_s A_{lm} + a^2\omega^2 - 2am\omega\,\). We emphasize that the perturbation is analyzed in a basis of spin-weighted spheroidal harmonics. We will later explain how the result extends to an initial perturbation expressed in terms of spherical harmonics, which is the most commonly studied case~\cite{Burko:1997tb,Gleiser:2007ti,Burko:2007ju,Burko:2013bra}.

As in the spherically symmetric case, we aim to construct the Green function associated with Eq.~\eqref{eq:teuk_radial}, which requires two linearly independent solutions of Eq.~\eqref{eq:teuk_radial}. In analogy with the Schwarzschild case, we will analyze Eqs.~\eqref{eq:teuk_radial} and ~\eqref{eq:teuk_angular} in the low-frequency limit. Equation~\ref{eq:teuk_angular} is particularly suitable for such an expansion. In this regime, one finds~\cite{Berti:2005gp}
\begin{equation}
\lambda = l(l+1) - s(s+1) - \frac{2ms^2}{l(l+1)} a\omega + \mathcal{O}((a\omega)^2)\,.
\end{equation}
Defining $F(r)=\Delta/(r^2+a^2)$, we can recast Eq.~\eqref{eq:teuk_radial} as
\begin{equation}\label{Kerr_waveeq}
F(r)^2 {}_s R_{lm}''(r) + F'(r) F(r) {}_s R_{lm}'(r) + V\, {}_s R_{lm}(r)=0\,,
\end{equation}
which has a form similar to Eq.~\eqref{waveq_r} and the following effective potential:
\begin{align}
V(r) = &\frac{K^2 - 2i s (r - M) K K }{(r^2 + a^2)^2}+ \notag\\\quad \quad &\frac{\Delta(r)\left(4i s \omega r - \lambda\right)}{(r^2 + a^2)^2}- G(r)^2 - f(r) G'(r)\,,
\end{align}
with 
\begin{equation}
G(r) = \frac{s(r - M)}{r^2 + a^2} + \frac{r\, \Delta(r)}{(r^2 + a^2)^2}\,.
\end{equation}
Following the generalization previously performed in the static case, we may analyze Eq.~\eqref{Kerr_waveeq} by considering a more general potential
\begin{equation}\label{VmodKerr}
    V(r) \to V(r)+ g(r)\,,
\end{equation}
with $g(r)$ satisfying condition~\eqref{cond}. In general, we notice that wave solutions at the horizon must take the form $e^{\pm i \tilde{\omega}r_*}$ where $r_*$ is defined via $dr_*/dr=F(r)$ and $\tilde{\omega} = \omega - m \Omega$ where $\Omega = \frac{a}{2 M r_+}$. By setting $_{s}R_{lm}(r)=\left(F(r)\right)^{-2IM\tilde{\omega}} \,_{s}w_{lm}(r)$ and introducing the dimensionless variable \(z = r\omega\), at leading order in $M\omega$ Eq.~\eqref{eq:teuk_radial} becomes
\begin{equation}\label{Kerr_smallfreq}
_{s}w_{lm}''(z) + \left(1 + \frac{2is}{z} - \frac{l(l+1)}{z^2} \right) \, _{s}w_{lm} + \mathcal{O}(M\omega) =0\,.
\end{equation}
The structure here slightly differs from the Schwarzschild case, since a term $ 2is /z$ is present in this formulation of the potential. Two linearly independent solutions of Eq.~\eqref{Kerr_smallfreq} are given by
\begin{align}
 & w_1 = z^{l+1}e^{iz}\Phi(l+s+1,2l+2,-2iz) \,,\notag\\ &  w_2 = z^{l+1}e^{iz}U(l+s+1,2l+2,-2iz)\,,
\end{align} where\footnote{{Although \( \Phi(a,b,z) \) is commonly referred to as \( M(a,b,z) \), we choose to use the notation \( \Phi \) in order to avoid confusion with the total mass \( M \) of the object.
}} \(\Phi(a,b,z)\) and \(U(a,b,z)\) are confluent hypergeometric functions. One can show that (see Appendix~\ref{KerrApp}):
\begin{itemize}
    \item \(\Psi_1=\left(F(r)\right)^{-2Mi\tilde{\omega}}w_1(r\omega)\) satisfies the  ingoing boundary condition at $r_+$.
    \item \(\Psi_2=\left(F(r)\right)^{-2Mi\tilde{\omega}}w_2(r\omega)\) describes an outgoing wave as \(r \to \infty\).
\end{itemize}
Moreover, the Wronskian of the two solutions is \(
W(\Psi_1, \Psi_2) = i(-1)^{l+1} \frac{(2l + 1)! \, \omega}{2^{2l+1} (l + s)!}+\mathcal{O}\left((M\omega)^2\right)\).
Following the same procedure adopted in the Schwarzschild case, the branch cut contribution to the Green function is given by
\begin{equation}
G_{\rm B}(r, r', \omega) =
\left[
\frac{\Psi_2(r, \omega e^{2\pi i})}{W(\omega e^{2\pi i})}
-
\frac{\Psi_2(r, \omega)}{W(\omega)}
\right]
\Psi_1(r', \omega)\,,
\end{equation}
and, using the results from Appendix~\ref{KerrApp}, we find that any potential of the type presented in Eq.~\eqref{VmodKerr} leads to
\begin{equation}
G_{\rm B}(r, r', \omega)=
\frac{(-1)^{l-s} 4\pi M \omega (l - s)!}{(2l + 1)!}
\frac{\Psi_1(r, \omega) \, \Psi_1(r', \omega)}{W(\omega)}\,.
\end{equation}
Analyzing the behavior of \(\Psi_1\), we find that for finite \(r\), corresponding to \(z = \omega r \to 0\) as \(\omega \to 0\), the leading order term in the small-frequency expansion is
\begin{equation}\label{Psi1z0}
\Psi_1 \simeq (\omega r)^{l+1} + \mathcal{O}((M\omega)^{l+2})\,,
\end{equation}
while, in the \(z \to \infty\) limit, the asymptotic behavior is
\begin{equation}\label{Psi1zinf}
\Psi_1 \simeq e^{i \omega r} (2l + 1)! \, e^{-i \pi (l + s + 1)/2} \, (2\omega)^{-(l + s + 1)} \, {r^{-s}\over (l - s)!}\,.
\end{equation}
Proceeding as in the Schwarzschild case, using Eqs.~\eqref{Psi1z0} and ~\eqref{Psi1zinf}, at {\it  timelike infinity} the Green function reads
\begin{equation}\label{Greeni+Kerr}
G_{B,i+}(r_*, t \mid r'_*, t')\sim  M C_l (rr')^{l+1} (t - t')^{-2l-3}\,,
\end{equation}
with $C_l= (-1)^{l + s+1} \, 2^{2l + 1} \, (2l+ 2)! \, (l + s)! \, (l - s)! /( \, [(2l + 1)!]^2)$. This coincides with the results found in Ref.~\cite{Hod:2000ya} in the case of a perturbation decomposed in spheroidal spin-weighted harmonics; here we showed that Eq.~\eqref{Greeni+Kerr} is valid also when modifying the Kerr potential as in Eq.~\eqref{VmodKerr}. Although not explicitly shown, also at {\it future null infinity} we find that results of Ref.~\cite{Hod:2000ya} are valid for the potential in Eq.~\eqref{VmodKerr}. 

We stress that here we analyzed how the perturbation behaves when considered in a decomposition in spheroidal spin weighted harmonics. Clearly, if the initial perturbation is considered in a decomposition in spin-weighted spherical harmonics, different modes will mix because of the rotating background. Indeed, for small $a\omega$~\cite{Hod:1999rx, Hod:2000fh,Berti:2005gp}
\begin{equation}\label{eq:S_expansion}
{}_sS_{lm}(\theta, a\omega) e^{im\phi} = \sum_{k} C_{lk}(a\omega)^{|l - k|} {}_sY_{km}(\theta), 
\end{equation}
where the coefficients $C_{lk}$ are $\omega$-independent. The key point is that, as long as the individual modes remain unmodified, their mixing proceeds as in the standard case (see, e.g.,~\cite{Burko:2013bra}), and the overall outcome is insensitive to potential modifications like those considered above. This result is in excellent agreement with the findings of~\cite{Poisson:2002jz}, where the propagation of a scalar field on a stationary, asymptotically flat background is studied without assuming any additional symmetries. That work demonstrates that the late-time tail behavior is insensitive to the specific symmetries of the background and is instead entirely governed by the multipolar structure of the perturbation. Our findings confirm this conclusion: differences in the power-law decay between the rotating and non-rotating cases arise solely from the distinct harmonic decompositions of the scalar field in the two geometries.


\section{Applications}

\subsection{Implications for the BH case}\label{bhcase}
\begin{figure}
    \centering \includegraphics[width=1.1\linewidth]{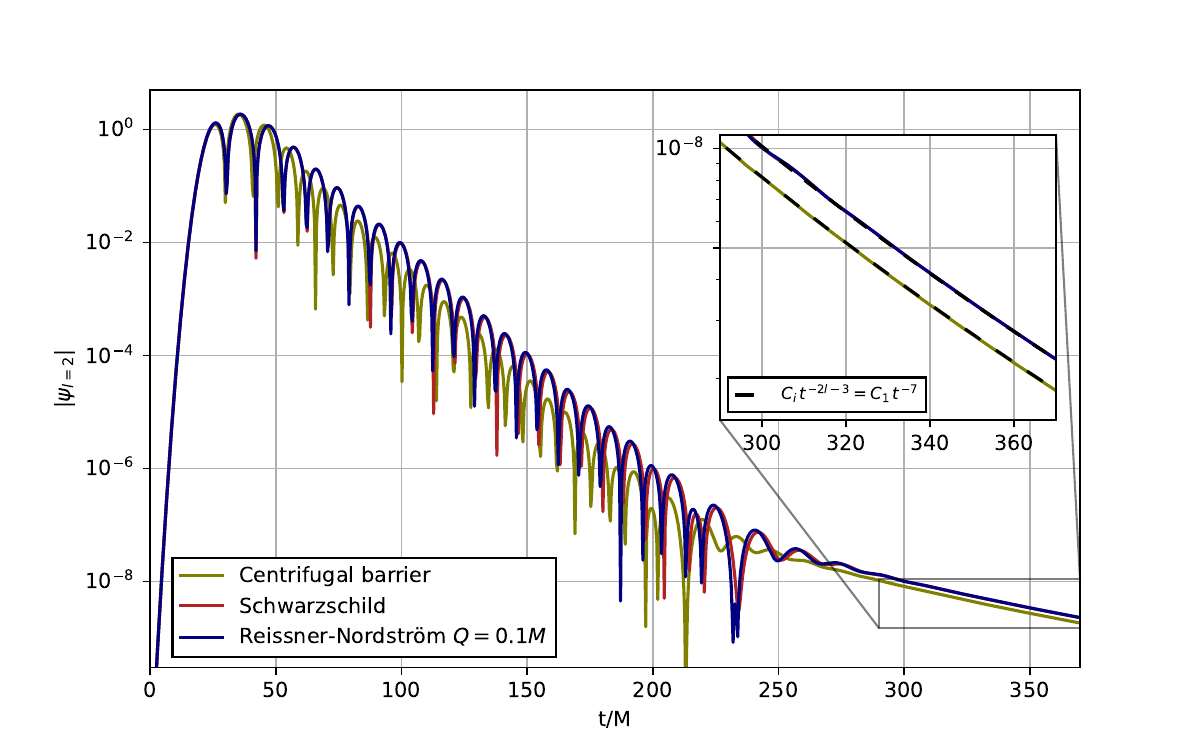}
    \includegraphics[width=1.1\linewidth]{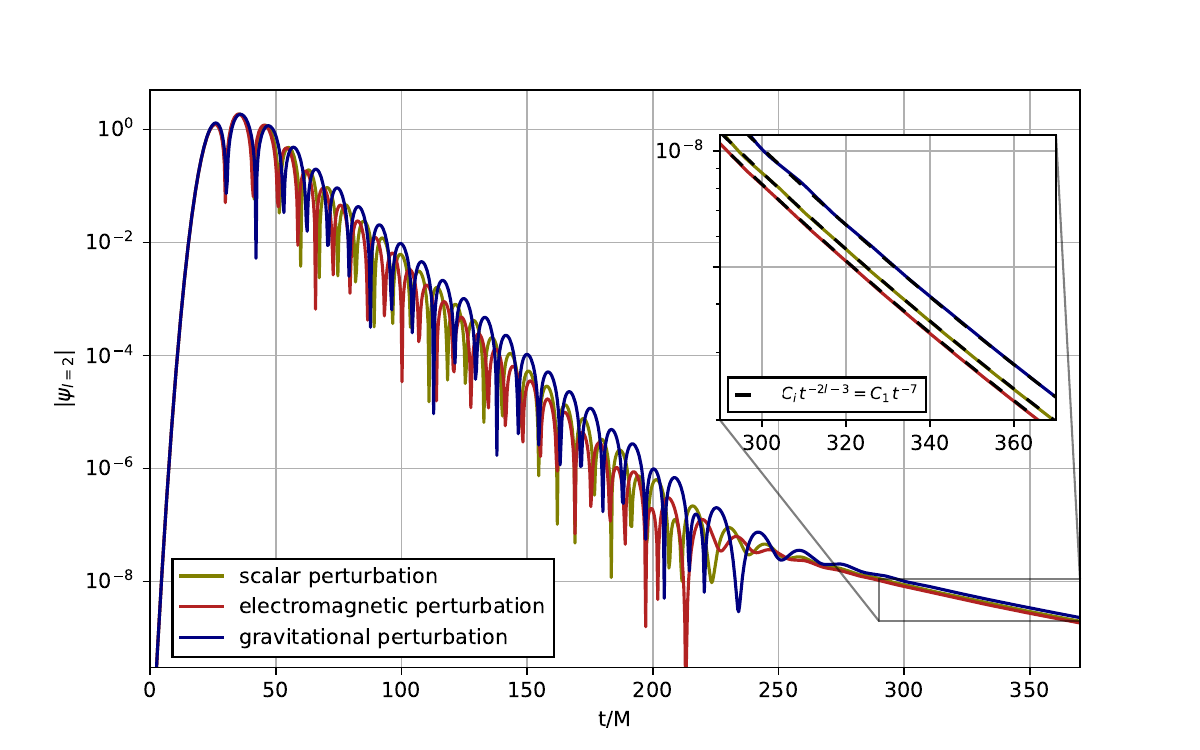}
    \includegraphics[width=1.1\linewidth]{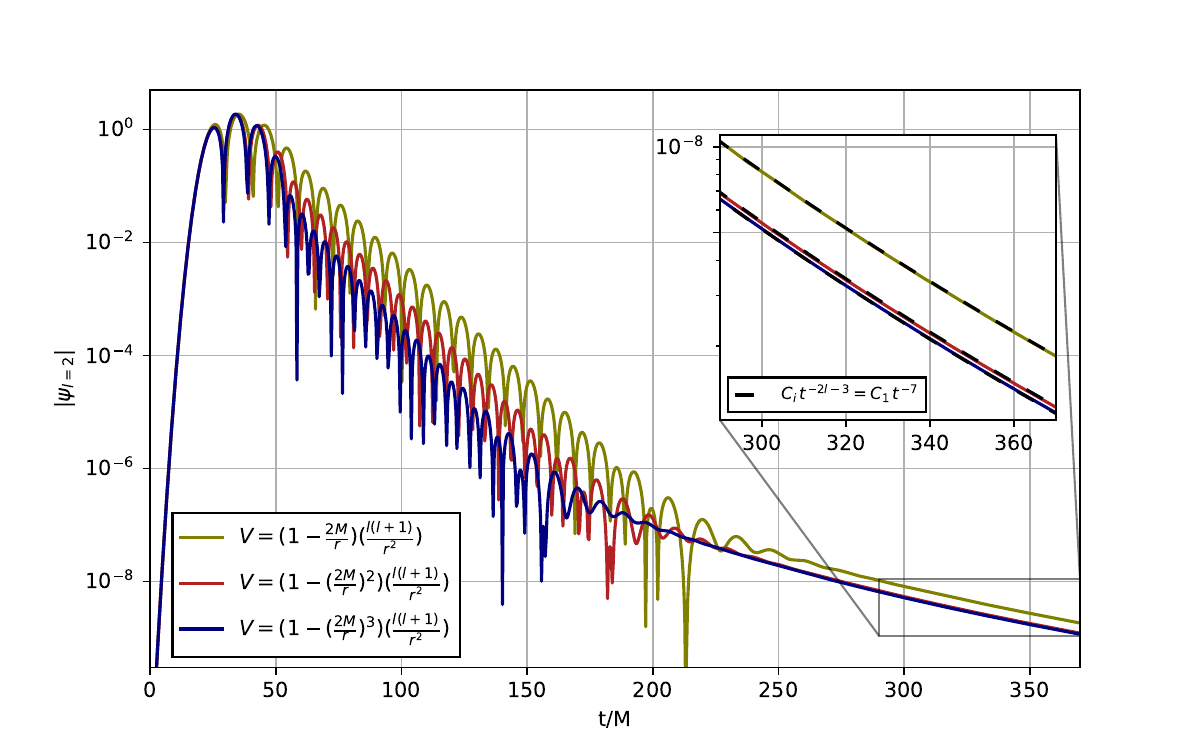}
    \caption{{\bf Upper panel}: evolution of a Gaussian packet for different configurations of Eq.~\eqref{waveeq} (See Appendix~\ref{Numerics}). The (2,2) waveform multipole is showed for: (1) the solely presence of a centrifugal barrier, namely $V(r)=f(r)l(l+1)/r^2$; (2) a gravitational perturbation on a  Schwarzschild background; (3)  a gravitational perturbation on a  Reissner–Nordström background. In all of these cases, the tail always scales as $t^{-2l-3}$. 
   {\bf Middle panel}: evolution of a Gaussian packet for perturbations of different spins on a Schwarzschild background (See Appendix~\ref{Numerics}). The (2,2) waveform multipole is showed for $|s|=0,1,2$. In all of these cases, the tail always scales as $t^{-2l-3}$. 
   {\bf Lower panel}: evolution of a Gaussian packet for different effective potentials in Eq.~\eqref{waveeq} (See Appendix~\ref{Numerics}). The (2,2) waveform multipole is showed for the solely presence of a centrifugal barrier, multiplied by a function $g(r)=1-(2M/r)^n$ with $n\geq1$. In all of these cases, the tail always scales as $t^{-2l-3}$.
   }
   \label{fig:perturb_tails}
\end{figure}

In the previous section we showed that the late-time tails in the linear response of a BH are not modified by the addition of a term of the form $1/r^\alpha$ to the effective potential, provided $\alpha > 2$. In particular, this result implies that:
\begin{itemize}
    \item Tails in Schwarzschild and Reissner--Nordström geometries coincide. This comparison with the purely centrifugal potential is shown in Fig.~\ref{fig:perturb_tails} (upper panel).
    
    \item Scalar, electromagnetic, and gravitational perturbations give rise to the same late-time tails. The Schwarzschild case is illustrated in Fig.~\ref{fig:perturb_tails} (middle panel).
    
    \item The addition of any correction of the form $1/r^\alpha$ with $\alpha > 2$ does not modify the tail behavior. Several examples are shown in Fig.~\ref{fig:perturb_tails} (lower panel).

    \item Tails of perturbations decomposed in spheroidal harmonics are insensitive to the BH spin.
\end{itemize}

All of the previous cases yield the same late-time tail at leading order in \( t \), scaling as \( \sim t^{-2l-3} \) at timelike infinity. The figures display the evolution of a Gaussian wave packet, computed using the numerical methods described in Appendix~\ref{Numerics}. The late-time tails showed in figure are extracted at timelike infinity.

\begin{figure}
    \centering
    \includegraphics[width=\linewidth]{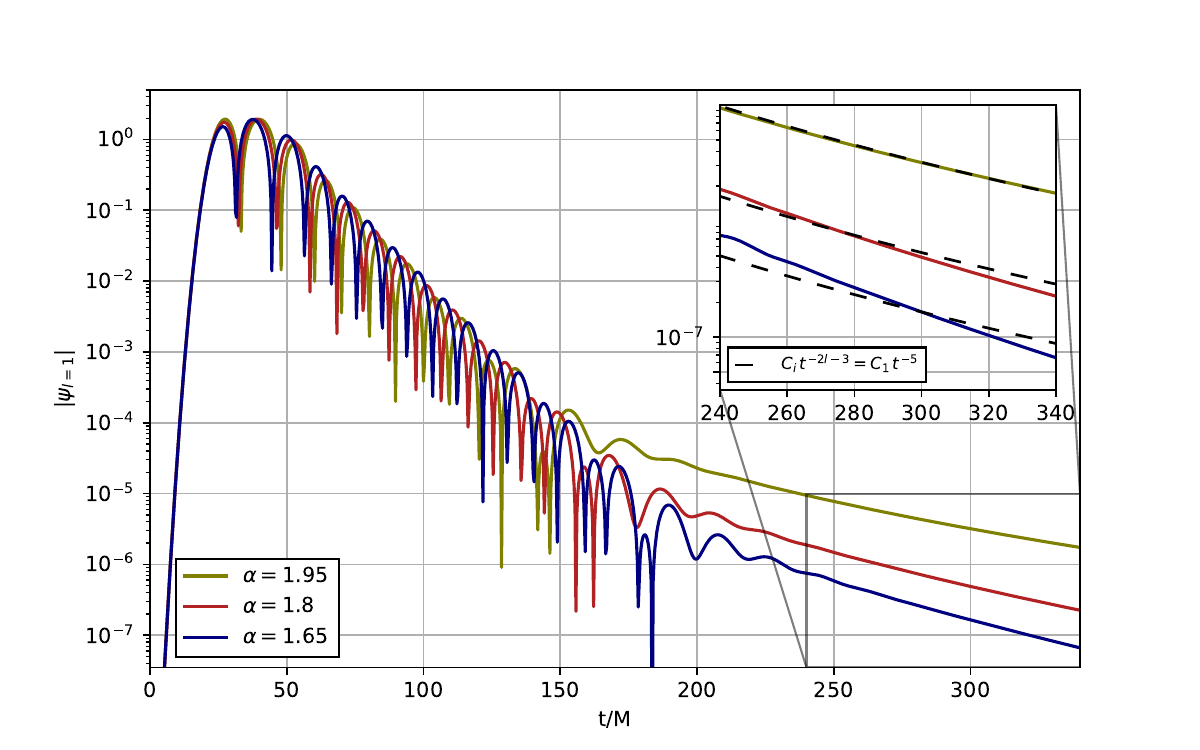}
    \includegraphics[width=\linewidth]{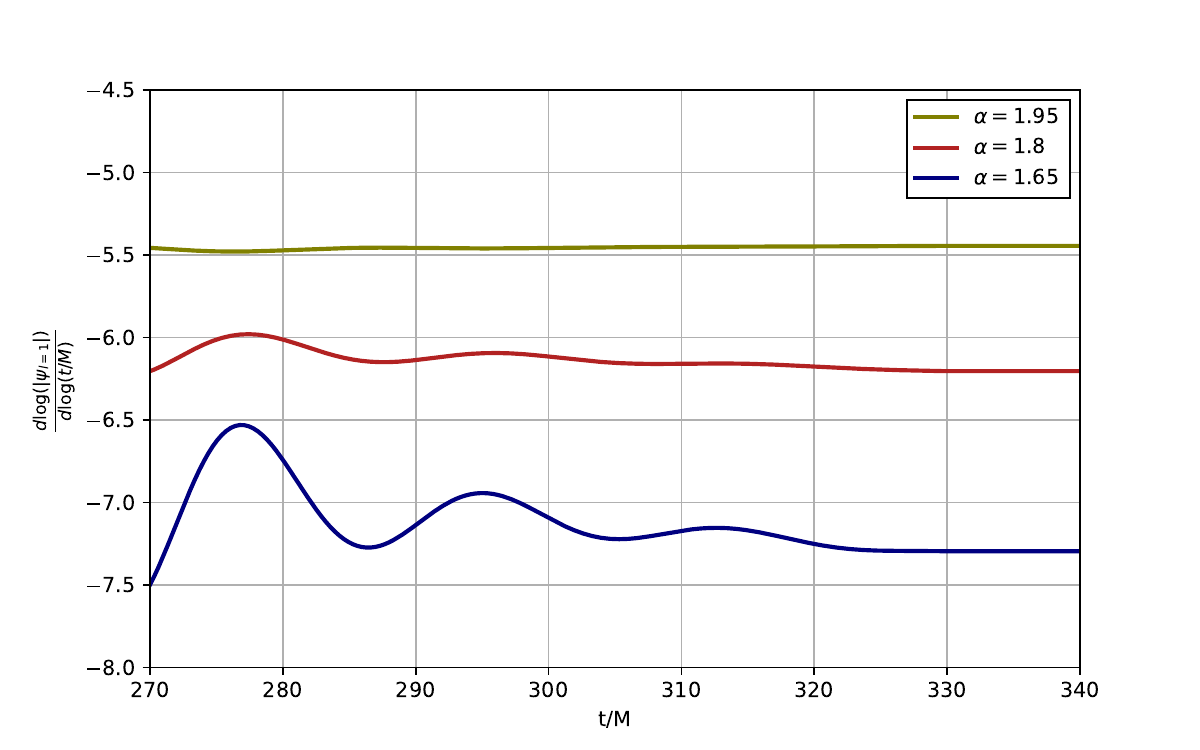}
    \caption{ Evolution of a Gaussian wave packet under different effective potentials as in Eq.~\eqref{waveeq} (see Appendix~\ref{Numerics}). In the top panel, the $(1,1)$ multipole of the waveform is shown. We modify the potential by adding a term of the form $V(r) = M^{\alpha-2}/r^\alpha$ with $1 \leq \alpha \leq 2$, in order to investigate how the late-time tail is affected. The farther $\alpha$ is from 2, the more the tail's power-law decay deviates from the standard $t^{-2l-3}$ behavior. For instance, when $\alpha = 1.95$, the behavior remains close to the centrifugal case, since $\alpha \sim 2$ effectively shifts $l(l+1) \to l(l+1)+1$, which for $l = 1$ changes the decay exponent from $-5$ to approximately $-5.4$. In the bottom panel, we show the local power index $d \log(|\psi|)/d \log(t/M)$ for the previous cases, as defined in \cite{Zenginoglu:2009ey}. It is evident that in all cases the decay rate differs from the canonical $-2l-3 = -5$ associated with a pure centrifugal barrier. Moreover, we observe that, for smaller values of $\alpha$, the ringdown phase persists longer.}
    \label{fig:variousalpha}
\end{figure}

It is now interesting to discuss the cases $\alpha = 2$ and $1 < \alpha < 2$. When $\alpha = 2$ (or analogously $\beta = 2$), the centrifugal term is clearly modified. In particular,
\begin{equation}
    \frac{l(l+1)}{r^2} \to \frac{l(l+1)+A}{r^2} = \frac{l'(l'+1)}{r^2}\,,
\end{equation}
where 
\begin{equation}
    l' = \frac{1}{2} \left( \sqrt{4A + 4l^2 + 4l + 1} - 1 \right)\,.
\end{equation}
Hence, the tail is modified as a consequence of the shift in the effective angular momentum. In particular, the decay rates in the Green function at timelike and future null infinity are given by $t^{-2l'-3}$ and $u^{-l'-2}$, respectively\footnote{Where $u=t-r_*$ is the retarded time.}. This result is consistent with the findings of \cite{Thomopoulos:2025nuf}, where modifications to the centrifugal term were shown to lead to different power-law behaviors for the late-time tail, impacting on the accuracy of ringdown waveform models.
On the other hand, the case $1 < \alpha < 2$ is more challenging to discuss analytically. It is certain that the tail gets modified, as non-vanishing $\omega$-dependent corrections are introduced into Eq.~\eqref{eqZpsi}. However, no analytical solution to Eq.~\eqref{eqZpsi} is known for a generic real value of $\alpha$ in the discussed range. Some illustrative examples are shown in Fig.~\ref{fig:variousalpha}, where clear deviations from the standard universal power-law scaling are evident.

Some examples of potentials of the form given in Eq.~\eqref{mod_pot} have been numerically investigated in~\cite{deMedeiros:2025ayq} for various values of~$\alpha$, and their results show perfect agreement with ours.

\subsection{BHs surrounded by matter profiles}\label{dmcase}
Consider a BH surrounded by some matter profile, modelled as a perfect fluid. The metric clearly differs from the vacuum case. Focusing on spherically symmetric distributions of matter on Schwarzschild-like backgrounds, the metric takes the generic form
\begin{equation}
    g^{(0)}_{\mu\nu} dx^\mu dx^\nu = -e^\nu dt^2 + e^\lambda dr^2 + r^2 d\Omega^2\,,
\end{equation}
and satisfies the Tolman-Oppenheimer-Volkoff equations
\begin{align}\label{TOV}
    & M' = 4\pi r^2 \rho, \quad \nu' = 2 \frac{M + 4\pi r^3 p}{r(r - 2M)}, \notag\\
    &\quad\quad p' = - (p + \rho) \frac{M + 4\pi r^3 p}{r(r - 2M)}\,,
\end{align}
where $p(r)$ and $\rho(r)$ correspond to the pressure and energy density of the matter profile. For gravitational axial perturbations, the linearized equations reduce to (see, e.g.,~\cite{Pani:2018inf})
\begin{align}\label{pert_matter}
    & e^{(\nu - \lambda)/2} \left( e^{(\nu - \lambda)/2} \psi' \right)' + \omega^2 \psi \notag\\
    &\quad\quad - e^\nu \left( \frac{l(l+1)}{r^2} - \frac{6M}{r^3} + 4\pi(\rho - p) \right) \psi = 0\,,
\end{align}
where $M = M(r)=M_{BH}+m(r)$ with $M_{BH}$ the mass of the central object. As proved in Sec.~\ref{generalization}, as long as the potential is asymptotically modified by terms of the form $1/r^\alpha$ with $\alpha > 2$, the late-time tails remain unaffected. Suppose that, asymptotically for $r \to \infty$,
\begin{equation}\label{rho}
    \rho \to \frac{C}{r^{\gamma}}\,,
\end{equation}
where $C$ carries the appropriate mass dimensions. One can asymptotically solve Eq.~\eqref{TOV} to investigate whether the presence of a matter profile could affect the tail behavior. From the mass equation, it follows straightforwardly that
\begin{equation}\label{m_asymp}
    m(r) \sim \frac{4\pi C}{3-\gamma} r^{3-\gamma}\,,
\end{equation}
which is physically meaningful if and only if $\gamma > 3$, as otherwise $m(r)$ would diverge at infinity. 
To solve the pressure equation, one may adopt the asymptotic ansatz $p \sim B/r^{\beta}$. Proceeding in this way, one finds that the only physical solution for $p$ is\footnote{In principle, the equation also admits the solution $p \sim -1/(4\pi r^2)$. However, for any physically reasonable value of the exponent $\gamma$ in Eq.~\eqref{rho}, this would lead to a divergent sound speed, rendering the solution unphysical.}

\begin{equation}\label{p_asymp}
    p \sim \frac{4 \pi  C^2 r^{2-2 \gamma}}{(\gamma -3)(2 \gamma -2)}\,.
\end{equation}
Finally, the equation for $\nu$ admits the asymptotic solution
\begin{equation}\label{nu_asympt}
 \nu \sim    \frac{8 \pi  C r^{2-\gamma }}{(\gamma
   -3) (\gamma -2)}\,.
\end{equation}
The above scalings show that in Eq.~\eqref{pert_matter} the potential is only modified by additive terms that do not affect the late-time tails. The same conclusion applies to the metric functions $\nu(r)$ and $\lambda(r)$.
Therefore, we demonstrated that modifications to the asymptotic structure of spacetime induced by matter profiles do not alter the universal late-time tail behavior.

\begin{figure}
    \centering
\includegraphics[width=\linewidth]{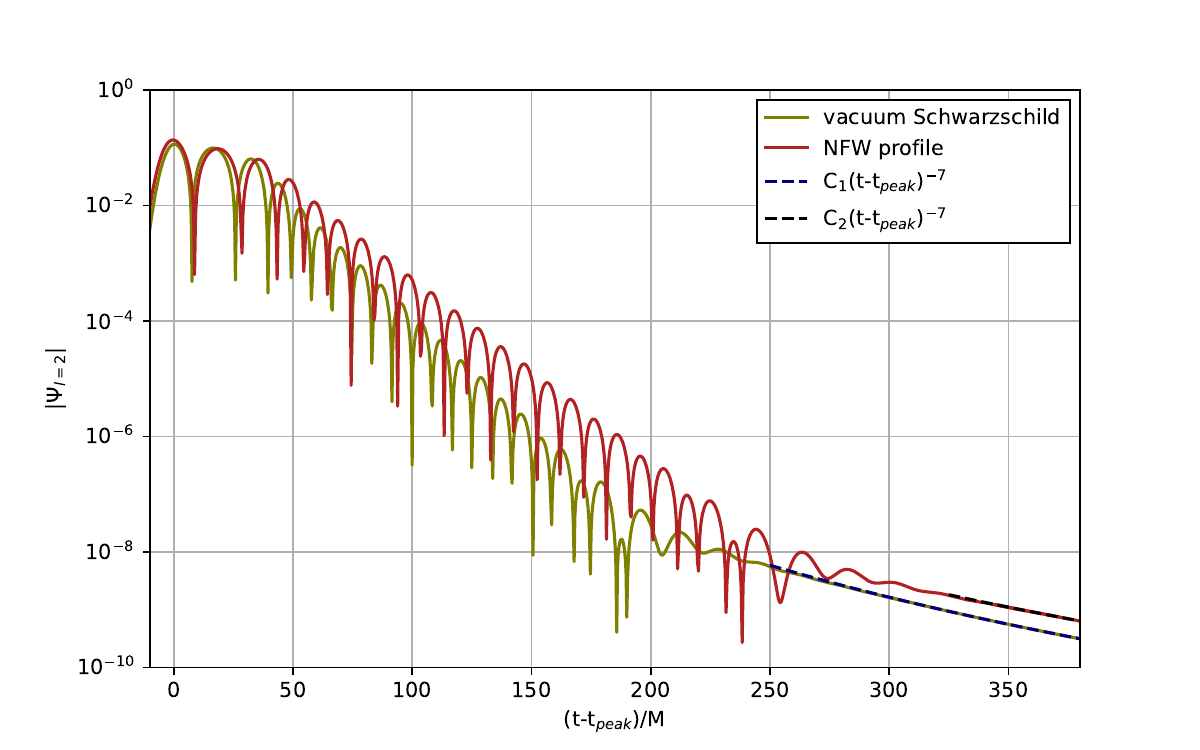}
    \caption{Evolution of a Gaussian packet for different effective potentials in Eq.~\eqref{waveeq} (See Appendix~\ref{Numerics}). The (2,2) waveform multipole is showed for the vacuum Schwarzschild solution and for Schwarzschild surrounded by a NFW profile~\cite{Navarro_1997}. We set $r_0=10^9 M$ and $r_1=10^8M$. Even though the ringdown is modified, the tails follow the same universal power law ($t^{-2l-3}=t^{-7}$, see dashed dark lines). We use the same initial condition in all  cases. The choice of $r_1$ is dictated by the fact that for our galaxy, by assuming a NFW profile, one has~\cite{Lin:2019yux} $r_1 \sim 6 \cdot 10^{8} M_{\text{SgrA}^*}$, where $ M_{\text{SgrA}^*}$ is the mass of Sagittarius A$^*$. For our galaxy, $r_0 \sim 6  \cdot 10^{11}M_{\text{SgrA}^*}$, however we chose a lower value in order to magnify the possible effect of the halo on time evolution.}
    \label{fig:SchwDM}
\end{figure}

There exist physically motivated matter profiles for which the parameter $\gamma$ defined in Eq.~\eqref{rho} is not larger than three. An example is provided by the celebrate Navarro-Frenk-White (NFW) profile for dark-matter distributions in galactic haloes~\cite{Navarro_1997}. 
In this case the problem of the divergence of $m(r)$ can be resolved by considering a matter distribution with finite support. For numerical convenience, we can study the effect of a cold dark matter distribution following an NFW profile~\cite{Navarro_1997} on the late-time tails, neglecting the pressure contribution. The NFW profile reads
\begin{equation}
       \rho(r)= \frac{\rho_0}{\left( \frac{r}{r_0} \right)\left( 1 + \frac{r}{r_0} \right)^2}  \,,
\end{equation}
 where $r_1=8 \pi \rho_0 r_0^3$. In the approximation $p \approx 0$, following~\cite{Zhang:2021bdr,Xu:2018wow} we can consider the case a Schwarzschild-like space-time, described by
\begin{equation}
    ds^2=-g(r) dt^2+{1 \over g(r)} dr^2 +r^2 d\Omega^2\,.
\end{equation}
The gravitational axial perturbations are described by Eq.~\eqref{waveeq} with the tortoise coordinate defined as $dr_*/dr=g(r)$ and the effective potential~\cite{Zhang:2021bdr} 
\begin{equation}
    V(r)=g(r) \Bigg( g''(r)+{g'(r) \over r}+
   {2g(r)+l(l+1)-2 \over r^2}\Bigg)\,.
\end{equation}
For a NFW profile, the background metric function reads~\cite{Xu:2018wow}
\begin{equation}
       g(r) = \left( 1 + \frac{r}{r_0} \right)^{ - \frac{r_1}{r}} - \frac{2 M}{ r} \,.
    \end{equation}
One can verify that the only corrections vanish faster than $1/r^2$ as $r \to \infty$, and thus the late-time tails remain unaffected. This is numerically confirmed in Fig.~\ref{fig:SchwDM}
One finds that, for both considered profiles, the effective potential only differs by terms $1/r^\alpha$ with $\alpha\geq 3$ from the vacuum case. This implies that, even though the ringdown will certainly be modified~\cite{Pezzella:2024tkf,Konoplya:2021ube,Cardoso:2021wlq}, tails will follow the same power-law of the vacuum case. This is numerically verified in Fig.~\ref{fig:SchwDM}.
\newpage
\subsection{Tails for horizonless compact objects}
The results of the previous sections assumed only ingoing waves as $r_* \to -\infty$, consistent with the BH scenario. However, when considering other types of compact objects, the physical boundary condition at the surface of the object will, in general, not correspond to a purely ingoing wave.
Provided the object is sufficiently compact so that the potential near its radius becomes negligible, Eq.~\eqref{waveeq} admits plane-wave solutions at the boundary~\cite{Cardoso:2019rvt,Maggio:2020jml}. We emphasize that this condition is not satisfied in the case of  neutron stars, which are not compact enough. Tails in the response of neutron stars were studied numerically in Ref.~\cite{Bernuzzi:2008rq}. As we shall show, our analytical analysis here is consistent with the results of~\cite{Bernuzzi:2008rq},and extends the latter to the case of exotic, horizonless, ultracompact objects~\cite{Cardoso:2019rvt}. In the scenario just described, the boundary condition near the radius of the object is~\cite{Maggio:2020jml}
\begin{equation}
    \psi\to \psi_{r_-} + R_{\text{ECO}}\, \psi_{r_+}\,,
\end{equation}
where $\psi_{r_+}$ satisfies a purely outgoing condition at the object's surface. This function can be obtained by complex conjugation of $\psi_{r_-}(r,\omega)$ in Eq.~\eqref{eqGreen}. 
Using $\mathcal{W}(\psi_{r+},\psi_{\infty_+})=2i\omega A^*_{\rm out}(\omega)$, we get
\begin{align}
  &\mathcal{W}(\psi_{r_-}+R_{\rm ECO}\psi_{r+},\psi_{\infty_+})=\notag\\&\quad\quad\quad\quad\quad=2i\omega \left(R_{\rm ECO}A^*_{\rm out}(\omega)+A_{\rm in}(\omega)\right)\,.
\end{align}
Proceeding exactly as for the BH case, one gets
\begin{widetext}
\begin{align}\label{greenAinAoutECO}
    &G_B^{\rm ECO}(r_*, t \mid r'_*, t') = \frac{1}{2\pi } \int_0^{-i\infty} d\omega\,e^{-i\omega(t - t')} K(\omega)  \notag\\ &\quad\quad\Bigg[\frac{\left(\psi_{r_-}(r_*<, \omega)+R_{\rm ECO} \psi_{r_+}(r_*<, \omega)\right)\left(\psi_{r_-}(r_*> , \omega)+R_{\rm ECO}\psi_{r_+}(r_*> , \omega)\right)}{2i\omega \left(A_{\text{in}}(\omega) +R_{\rm ECO}A^*_{\text{out}}(\omega)\right)\left[A_{\text{in}}(\omega) + K(\omega) A_{\text{out}}(\omega)+R_{\rm ECO}\left(A^*_{\text{out}}(\omega) + K(\omega) A^*_{\text{in}}(\omega)\right)\right]}\Bigg] \,.
\end{align}
\end{widetext}
In the case $R_{\text{ECO}} = 0$, Eq.~\eqref{greenAinAout} recovers the BH case, as expected. Since we are interested in the leading-order tail, we can neglect the subleading terms in $M\omega$. Moreover, in the $\omega\to 0$ limit, 
\begin{equation}
    \lim_{\omega \to 0} A^*_{\text{out}}(\omega)=\lim_{\omega \to 0} A_{\text{in}}(\omega)\,,
\end{equation}
hence in this limit
\begin{align}
    &G_B^{\rm ECO}(r_*, t \mid r'_*, t') \sim \frac{1}{\left(1 +R_{\rm ECO}\right)^2}\left(G_B^{\rm BH}(r_*, t \mid r'_*, t')+\right.\notag \\&\quad  \left.R_{\rm ECO}\,G_{B,1}(r_*, t \mid r'_*, t')+R_{\rm ECO}^2\,G_{B,2}(r_*, t \mid r'_*, t')\right)\,,
\end{align}
with
\begin{align}
    &G_{B,1}(r_*, t \mid r'_*, t') = \frac{1}{2\pi } \int_0^{-i\infty} d\omega\,e^{-i\omega(t - t')} K(\omega)  \notag\\ &\Bigg[\frac{\psi_{r_-}(r_*<, \omega)\psi_{r_+}(r_*> , \omega)+\left(r_*<\iff r_*>\right)}{2i\omega A_{\text{in}}(\omega)^2 }\Bigg] \,,
\end{align}
and
\begin{align}
    &G_{B,2}(r_*, t \mid r'_*, t') = \frac{1}{2\pi } \int_0^{-i\infty} d\omega\,e^{-i\omega(t - t')} K(\omega)  \notag\\ &\quad\quad\Bigg[\frac{\psi_{r_+}(r_*<, \omega)\psi_{r_+}(r_*> , \omega)}{2i\omega A_{\text{in}}(\omega)^2 }\Bigg] \,.
\end{align}
Now, using the solutions for $M\omega \to 0$ we have 
\begin{align}
    G_{B,1}&(r_*, t  \mid r'_*, t')\sim 2iM \int_0^{-i\infty} d\omega\, \left(F^*_l(0, \omega r) F_l(0, \omega r')\right.\notag\\&\left.+F_l(0, \omega r) F^*_l(0, \omega r') \right)e^{-i\omega(t_r - t'_r)} \,,
\end{align}
and 
\begin{align}
    G_{B,2}&(r_*, t  \mid r'_*, t')\sim \notag\\&2iM \int_0^{-i\infty} d\omega\,F^*_l(0, \omega r) F^*_l(0, \omega r')e^{-i\omega(t_r - t'_r)} \,,
\end{align}
where $t_r,t_r'$ correspond to the retarded times defined in Eq.~\eqref{retardedtime}. Both at timelike and future null infinity, one obtains 
\begin{align}
   G_{B,1}(r_*, t  \mid r'_*, t')&=2G_{B}^{\rm BH}(r_*, t  \mid r'_*, t')\,, \\
   G_{B,2}(r_*, t  \mid r'_*, t')&=G_{B}^{\rm BH}(r_*, t  \mid r'_*, t')\,.
\end{align}
Therefore, the total branch-cut contribution to the ECO  Green function is the same as that of the BH,
\begin{equation}
    G_{B}^{\rm ECO}(r_*, t  \mid r'_*, t')=G_{B}^{\rm BH}(r_*, t  \mid r'_*, t')\,,
\end{equation}
and consequently the power-law tail is the same.
{Surprisingly, the contribution to the Green function arising from the wave component reflected at the surface of the object does not vanish. However, in the low-frequency limit, ingoing and outgoing plane waves near the would-be horizon become practically indistinguishable, contributing equally to the total Green function. Specifically, one finds that imposing purely ingoing or purely outgoing boundary conditions at the radius of the object yields the same exact result when computing the late-time Green function.}

\begin{figure}[H]
    \centering    \includegraphics[width=\linewidth]{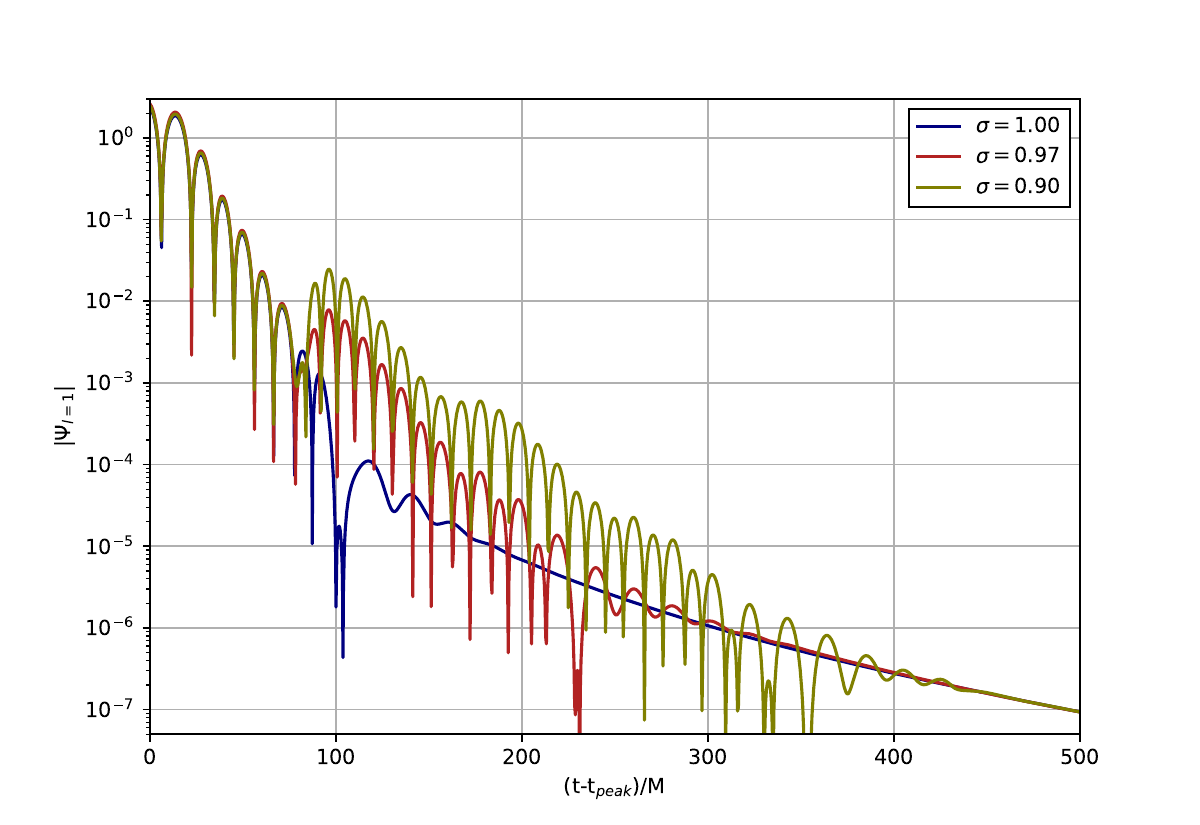}
    \caption{
    Comparison of the time evolution of a Gaussian wave-packet for a BH and for horizonless compact object. The parameters $\sigma$ quantifies the probability for a wave to be ingoing at the would-be horizon. In particular, $\sigma=1$ corresponds to the BH case. When $\sigma<1$, also reflection takes place at the object radius. One clearly sees that the BH case ($\sigma=1$) shows the  well-known late-time power law decay. In the case $\sigma=0.97$, an echo is clearly visible for times $t\sim 90M$, after which the perturbation will decay with the same power law tail, for times $t\sim 300M$. In the $\sigma=0.90$ one sees more echoes of the prompt signal, and a longer time is needed for the signal to show the power law decay.}
    \label{fig:ECOtails}
\end{figure} 

In Fig.~\ref{fig:ECOtails}, we show an example of the time evolution on the background of a toy model for an ultracompact horizonless object, obtained simply by modifying the boundary conditions near the would-be horizon of a Schwarzschild geometry.
Partial reflection at the radius is numerically implemented, as discussed in Appendix~\ref{Numerics}, via a parameter $\sigma$ that characterizes the ingoing or outgoing nature of the wave. 
Notably, for horizonless ultracompact objects, the ringdown is followed by echoes of the signal~\cite{Cardoso:2016oxy,Cardoso:2016rao,Cardoso:2017cqb} related to the absence of a horizon. However, at sufficiently late times, the signal will decay with the same power-law tail of BHs.

\subsection{Universal $1/t^2$ tails in the reflection amplitude}\label{Refl_sec}
Finally, in this section we discuss how late-time tails arise as a universal feature of BH scattering, particularly when considering the effective reflectivity of a BH. This aspect plays a crucial role in the study of reflectivity and greybody factors as gravitational-wave observables, as proposed in recent work~\cite{Oshita:2023cjz,Okabayashi:2024qbz,Rosato:2024arw,Rosato:2025byu}. 

Let $R(\omega)$ denote the reflection amplitude of the compact object under consideration, namely the relative amplitude of the reflected wave in a scattering process (see, e.g.,~\cite{Rosato:2024arw}) . From the behavior of $\psi_{r_-}(r,\omega)$ around the branch cut (see Appendix~\ref{branchcutapp}), it can be easily seen that
\begin{equation}
R(\omega e^{2\pi i})={A_{\rm out}(\omega e^{2\pi i}) \over A_{\rm in}(\omega e^{2\pi i}) }={A_{\rm out}(\omega)- K^*(\omega)A_{\rm in}(\omega)\over A_{\rm in}(\omega)+ K(\omega)A_{\rm out
}(\omega)}\,.
\end{equation}
When accounting for the branch cut contribution in the Fourier transform of $R(\omega)$, we have
\begin{align}
    &\mathcal{F}_{B}\left(R\right)(t)=-\int_{0}^{-i\infty}{d\omega\over 2\pi}\, e^{-i\omega t} \left(R(\omega e^{2\pi i})-R(\omega)\right)=\notag\\& \int_{0}^{-i\infty}{d\omega\over 2\pi} \,e^{-i\omega t}  \left({K^*(\omega)A^2_{\rm in}(\omega)+K(\omega)A^2_{\rm out}(\omega) \over A^2_{\rm in}(\omega)+K(\omega)A_{\rm out
}(\omega) A_{\rm in}(\omega)}\right)\,.
\end{align}
Again, in the $t \to \infty$ limit, the integral is dominated by the contribution from $\omega \to 0$. Using the small-frequency behavior derived in Sec.~\ref{greenfunctiondiscussion}, one finds, at leading order in $\omega$,
\begin{align}
    \mathcal{F}_{B}\left(R\right)(t)\sim (-1)^{l+1}4 M \int_{0}^{-i\infty}d\omega\, \omega e^{-i\omega t}=\notag\\(-1)^{l+1}4 M t^{-2}\quad\,.
\end{align}
Moreover, from the results of Ref.~\cite{Rosato:2025byu} (specifically Eq.~(B21)), one can see that for small frequencies $R_{\text{ECO}} = R_{\text{BH}}$.\footnote{Physically, this is due to the fact that a wave originating from infinity at very small frequencies is completely reflected by the effective gravitational barrier, without interacting with the near-horizon (or would-be horizon) structure.} 
From this, it follows that our results also generalize to exotic compact objects. 

\begin{figure}[H]
    \centering
    \includegraphics[width=\linewidth]{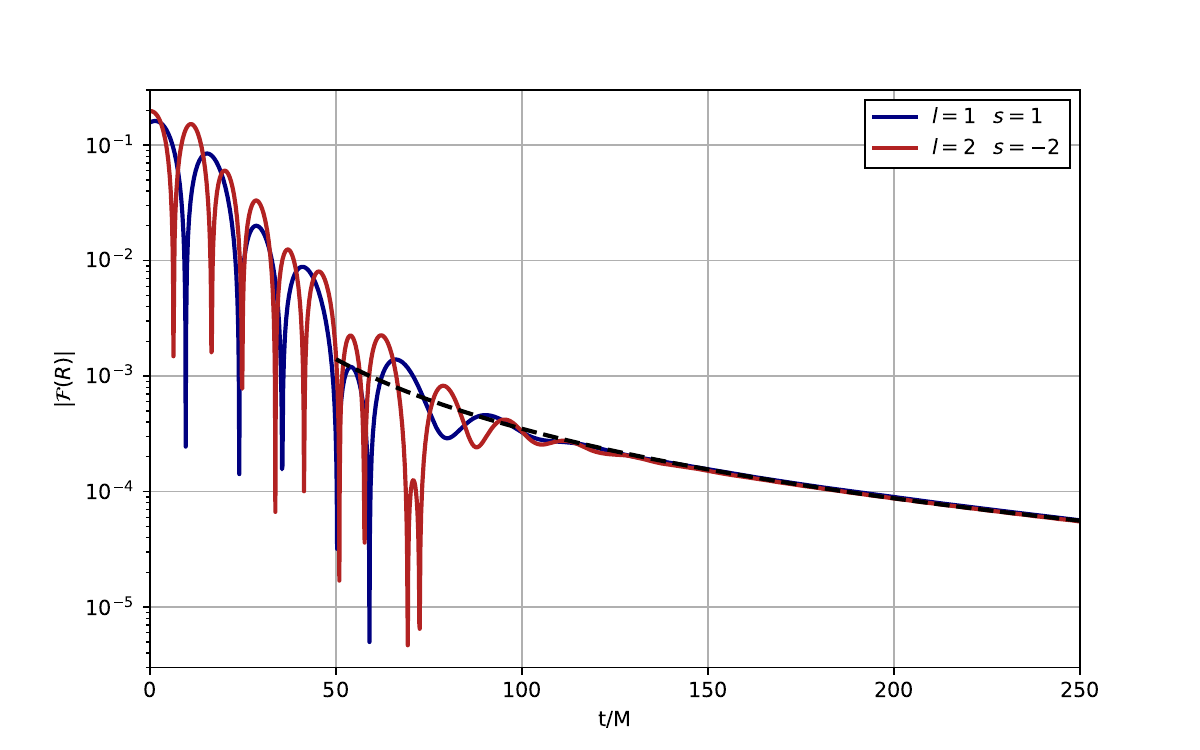}
    \caption{Time-domain reflectivity for different BH perturbations: the $l=1$ case for an electromagnetic perturbation and the $l=2$ case for a gravitational perturbations. In both cases the tail coincides withe the theoretical result, $t^{-2}$ (black dashed line).}
    \label{fig:refl}
\end{figure}

Therefore, we find that the Fourier transform of the reflectivity exhibits a \emph{universal late-time tail, also independent of the angular number $l$}. This behavior has also been verified numerically. Different cases are shown in Fig.~\ref{fig:refl}, in particular the $l=1$ case for an electromagnetic perturbation and the $l=2$ case for a gravitational perturbation. In both cases, the tail shows excellent agreement with the theoretical prediction.

\section{Conclusions}
We have presented a comprehensive analytical and numerical investigation of the universality of late-time power-law tails in the linear response of compact objects to perturbations. By analyzing the low-frequency behavior of the Green function associated with a broad class of effective potentials, we have shown that the asymptotic decay rate of the tail is universaly determined by the large-distance structure of the effective potential in the Schwarzschild radial coordinate.

Our analysis demonstrates that tails are universally governed by a $t^{-2l-3}$ decay law at timelike infinity for all  configurations in which the effective potential asymptotically approaches a centrifugal barrier with subleading corrections falling off faster than $1/r^2$. This result encompasses vacuum spacetimes such as Schwarzschild and Reissner–Nordstr\"om, matter profiles around BHs, perturbations of spinning BHs described by the Teukolsky equation, and even horizonless compact objects with partially reflective boundaries. We also extended and revised earlier analyses and provided an accurate characterization of scenarios in which deviations from the universal behavior can occur, namely for potentials with long-range corrections decaying more slowly than $1/r^2$.

We have further explored the implications of our findings for BHs surrounded by physically motivated matter distributions, including those following the NFW profile for dark matter. Our results show that the tails remain unaffected in such cases, reinforcing the robustness of the universal behavior. Similarly, we demonstrated that ultracompact, horizonless objects exhibit the same late-time tail structure as BHs, despite differing boundary conditions and the presence of echoes at earlier times in their linear response.

Finally, we discussed the emergence of tails in the context of the reflectivity and greybody factor, highlighting how the late-time decay is encoded in the branch-cut structure of the reflection amplitude. The resulting tail is found to be universal and decaying as $1/t^2$, independently of the multipole number.

Overall, our results establish a firm theoretical foundation for understanding the origin and universality of gravitational-wave tails. They also offer practical tools for interpreting late-time features in numerical simulations and future observations, especially in the era of precision gravitational-wave astronomy.

Finally, we remark that our analysis captures the very late-time regime, as we focus solely on the dominant power-law decay of the tail. Further investigation is required at intermediate times, where the behavior of the signal can deviate from the asymptotic regime and can depend on the inspiral dynamics and initial conditions of the system~\cite{DeAmicis:2024not}. In particular, when considering non-Schwarzschild backgrounds, the dynamics might be modified, and a careful analysis will therefore be required to model the tails in the intermediate-time regime.

\begin{acknowledgments}
We thank Gregorio Carullo and Marina de Amicis for fruitful discussions and comments on the draft. 
This work is partially supported by the MUR PRIN Grant 2020KR4KN2 ``String Theory as a bridge between Gauge Theories and Quantum Gravity'' and by the FARE programme (GW-NEXT, CUP:~B84I20000100001).     
\end{acknowledgments}

\appendix 
\section{Relevant formulas}\label{connectionformulas}
In this section, we recall some mathematical results that are useful throughout the paper. In particular, the analytic solutions of the perturbation equations in the main text are expressed via Coulomb wave functions and Kummer functions. Coulomb wave functions are solutions to the following differential equation 
\begin{equation}\label{CoulombEq}
    {d^2 u_l(\eta,z) \over dz^2}+\left(1-{2\eta \over z}+{l(l+1)\over z^2}\right)u_l(\eta,z)=0\,,
\end{equation}
where we can express $u_l(\eta,z)$ as
\begin{equation}
    u_l(\eta,z)= c_1 F_l(\eta,z)+c_2 G_l(\eta,z)\,.
\end{equation}
Eq.~\eqref{CoulombEq} can be connected to Kummer equation
\begin{equation}\label{KummerEq}
z \frac{d^2 w}{dz^2} + (b - z) \frac{dw}{dz} - a w = 0\,,
\end{equation}
the solution of which can be expressed as a linear combination of confluent hypergeometric functions
\begin{equation}
    w(a,b,z)=c_1 \Phi(a,b,z)+c_2 U(a,b,z)\,.
\end{equation}
There exist interesting connection formulas between solutions to Eq.~\eqref{CoulombEq} and Eq.~\eqref{KummerEq}, in particular it holds that~\cite{NIST:DLMF,Gaspard:2018xgb}
\begin{equation}
    F_l(\eta, \rho) = C_l(\eta)\, \rho^{l + 1} e^{\mp i \rho} \Phi(l + 1 \mp i\eta, 2l + 2, \pm 2i\rho)\,,
\end{equation}
where
\begin{equation}
    C_l(\eta) = \frac{2^l e^{-\pi \eta / 2} \left| \Gamma(l + 1 + i\eta) \right|}{(2l + 1)!}\,.
\end{equation}
Moreover, if one defines
\begin{equation}
    H_l^{\pm}(\eta, \rho) = G_l(\eta, \rho) \pm i F_l(\eta, \rho)\,,
\end{equation}
it holds that
\begin{align}
   & e^{\mp i \theta_l(\eta, \rho)}H_l^{\pm}(\eta, \rho) = \notag\\& = (\mp 2i\rho)^{l + 1 \pm i\eta} U(l + 1 \pm i\eta, 2l + 2, \mp 2i\rho)\,,
\end{align}
with
\begin{equation}
    \theta_l(\eta, \rho) = \rho - \eta \ln(2\rho) - \frac{1}{2} l \pi + \text{ph} \, \Gamma(l + 1 + i\eta)\,.
\end{equation}
When considering a complex rotation of $e^{2\pi i}$ there are interesting properties one can take into account. In particular, $\Phi(a,b,z)$ and consequently $F_l(\eta,\rho)$ are {\it single-valued} functions. However, $U(a,b,z)$ and consequently $H_l^{\pm}(\eta, \rho)$ are {\it multi-valued functions}. In particular $U(a,b,z)$ is multi-valued with respect to $z$, for $b\in\mathbb{Z}$ it satisfies 
\begin{align}\label{branchUM}
U(a, b, z e^{2\pi i}) &= U(a, b, z) +\notag\\ &2\pi i \, \frac{(-1)^{b}}{(b-1)! \, \Gamma(a +1- b)} \, \Phi(a, b, z)\,.
\end{align}
A straightforward algebraic manipulation then leads to
\begin{align}\label{branchcut_Hpm}
    &H^{+}_l(\eta,\rho e^{2\pi i})=H^{+}_l(\eta,\rho)+ \notag\\&
\quad \quad-K_{l,\eta}e^{-i \eta \ln(2\rho)}\rho^{i\eta} \left(  H^{+}_l(\eta,\rho)-  H^{-}_l(\eta,\rho)\right)\,,
\end{align}
with
\begin{equation}
    K_{l}(\eta)=  {2\pi i (-1)^{l+i\eta}e^{-\pi {\eta \over 2}} \over i\Gamma(l+1-i\eta)\Gamma(-l+i\eta)}\,.
\end{equation}
In particular, one finds that
\begin{align}
    &\lim_{\eta \to 0^+}H^{+}_l(\eta,\rho e^{2\pi i})=H^{+}_l(\eta,\rho)+\notag\\&\quad+2\pi \eta \left(  H^{+}_l(\eta,\rho)-  H^{-}_l(\eta,\rho)\right)+\mathcal{O}(\eta^2)\,.
\end{align}

\section{Branch cut}\label{branchcutcomputation}

We are interested in computing how $\psi_{\infty_+}(r_*,\omega)$ changes under a full rotation around the origin in the complex $\omega$-plane. To this end, we consider Eq.~\eqref{waveeq} and adopt a different change of variables. In particular, we set
\begin{equation}
    \Psi_{\rm lm}={1 \over\sqrt{f(r)}}y_{\rm lm}(r)\,,
\end{equation}
and again solve in $z=\omega r$. Up to order $\mathcal{O}(\omega)$, the equation we obtain is
\begin{align}\label{eqorderomega}
  & y''(z)+ \left(1-\frac{l (l+1)}{z^2}\right) y(z) + M \omega\frac{4}{z}y(z)\notag\\&\quad -M \omega\frac{2 (l (l+1)-1)}{z^3}y(z)+\mathcal{O}\left((M\omega)^2\right)=0\,.
\end{align}
At leading order ($\mathcal{O}\left((M\omega)^0\right)$), the solution coincides with Eq.~\eqref{setofindependentsolutions}. In particular, we are interested in the ingoing and outgoing solutions at infinity, represented by
\begin{equation}
    \psi_{\infty \pm}={e^{ \pm i {\pi l \over 2}}\over\sqrt{f(r)}}H^{\pm}_{\rm l}(0,r\omega)
\end{equation}
where
\begin{equation}
    H^{\pm}_{\rm l}(\eta,z)=F_l(\eta,z)\pm iG_l(\eta,z)\,.
\end{equation}
Now, for $l \in \mathbb{Z}$, (see~\cite{Leaver:1986vnb}) 
\begin{align}
  &H^{\pm}_{\rm l}(\eta,z e^{2\pi i})= H^{\pm}_{\rm l}(\eta,z)+ \mathcal{O}(\eta)\,,
\end{align}
hence for $\eta=0$ there is no modification due to $z\to z e^{2\pi i}$, namely
\begin{equation}
    \psi_{\infty_+}(r,\omega e^{2\pi i})=\psi_{\infty_+}(r,\omega )+ \mathcal{O}\left(M\omega\right)\,.
\end{equation}
In order to compute the modification of the solution when the branch cut is crossed we have to solve Eq.~\eqref{eqorderomega} including $\mathcal{O}\left(M\omega\right)$. The latter is solvable via a series of Coulomb functions. Being $u_{l}(\eta,z)$ a generic Coulomb function, one gets
\begin{equation}
    {d^2 u_l(\eta,z) \over dz^2}+\left(1-{2\eta \over z}+{l(l+1)\over z^2}\right)u_l(\eta,z)=0\,.
\end{equation}
If we set $\eta=-2M\omega$, we can express the solution as 
\begin{equation}
    y_l(z)=\sum_{L} b_L u_{l+L}(\eta,z)\,,
\end{equation}
allowing for both positive and negative $L$. The Coulomb wave functions satisfy the recurrence relation
\begin{align}
\frac{1}{2L + 2l + 1} R_{L+1} u_{L+l+1} - \left( \frac{1}{z} + Q_L \right) u_{L+l}
 \notag\\+\frac{1}{2L + 2l + 1} R_L u_{L+l-1} = 0\,,
\end{align}
and the differential relation
\begin{align}
&\frac{d}{dz} u_{L+l} = -\frac{L + l}{2L + 2l + 1} R_{L+1} u_{L+l+1}\notag \\&\quad \quad- Q_L u_{L+l}
+ \frac{L + l + 1}{2L + 2l + 1} R_L u_{L+l-1}\,,
\end{align}
where
\begin{equation}
Q_L = \frac{\eta}{(L + l)(L + l + 1)}
\end{equation}
and
\begin{equation}
R_L = \frac{\left[ (L + l)^2 + \eta^2 \right]^{1/2}}{(L + l)}\,.
\end{equation}
By employing the aforementioned relations and substituting the series expansion into the equation, it follows that the coefficients \( b_L \) satisfy the following recurrence relation
\begin{equation}
    \alpha_L b_{L+1}+\beta_L b_L+ \gamma_L b_{L-1}=0\,,
\end{equation}
where
\begin{align}
    &\alpha_L=-M\omega {R_L \over 2L+2l-1}(2l(l+1)+2)\,, \\
    &\beta_L=(L+l)(L+l+1)-\notag\\&\quad \quad \quad \quad \quad l (l+1)+(2l(l+1)+2)Q_L \,,\\
    &\gamma_L=-M\omega {R_{L+1} \over 2L+2l+3}\left(2l(l+1)+2\right)\,.
\end{align}
One can check that the infinite series we have characterized is converging. However, since we are only interested in the solution up to order $\mathcal{O}\left(M\omega\right)$, by setting $b_0=1$, we easily find that $b_L$ is of order $\mathcal{O}\left((M\omega)^{|L|}\right)$, hence we we will consider only the terms $L=\pm 1,0$. For large values of $z$,
\begin{equation}
    H_{l}^{\pm}(\eta,z) \to e^{\pm i\left(z - \eta \ln(2z)-\phi_\pm\right)}\,,
\end{equation}
where 
\begin{equation}
    \phi_{\pm,l} =\pm \left(L+l\right){\pi \over 2}+{i \over 2} \ln\left({\Gamma(l+1- i\eta) \over \Gamma(l+1+ i\eta)}\right)\,.
\end{equation}
Clearly this implies that a total phase has to be reabsorbed in order to match the asymptotic behavior of $\psi_{\infty \pm}$. By reabsorbing the phase (see~\cite{Leaver:1986gd}) and using Eq.~\eqref{branchcut_Hpm}, one finds that in the limit \( M\omega \to 0 \)
\begin{align}\label{eqKvalue}
  &\psi_{\infty +}(r,\omega e^{2\pi i})=\psi_{\infty +}(r,\omega)-(-1)^l 2 \pi \eta\psi_{\infty -}(r,\omega)\notag\\&\quad = \psi_{\infty +}(r,\omega)-(-1)^l 4 M \pi \omega \,\psi_{\infty -}(r,\omega)+ \mathcal{O}\left((M\omega)^2\right)\,. 
\end{align}
It is now interesting to consider the problem when the modified potential in Eq.~\eqref{mod_pot} is taken into account. Without loss of generality, we can focus on the following potential 
\begin{equation} V(r) = f(r) \left( \frac{l(l+1)}{r^2} + \frac{A}{r^\alpha} \right)\,, \end{equation}
the generalization for  Eq.~\eqref{mod_pot} will be straightforward. We proceed exactly as before in the analysis of Eq.~\eqref{waveeq}, setting $\Psi_{lm} = y_{lm} / \sqrt{f(r)}$ and changing variables to $z = \omega r$. Expanding in powers of $M\omega$, we observe that Eq.~\eqref{eqorderomega} is modified depending on the value of $\alpha$: 
\begin{itemize}
\item For $\alpha > 3$, Eq.~\eqref{eqorderomega} receives corrections only at order $\mathcal{O}\left((M\omega)^{\alpha-2}\right)$ (since a factor of $\omega^2$ has been factored out in Eq.~\eqref{eqorderomega}), which vanishes faster than the linear term as $M\omega \to 0$. Therefore, Eq.~\eqref{kvalue} automatically remains valid at order $\mathcal{O}\left(M\omega\right)$, in particular the leading order value of $K(\omega)$ is untouched. \\
\item For $ 2 < \alpha < 3$, Eq.~\eqref{eqorderomega} would in principle receive corrections at order $\mathcal{O}\left((M\omega)^{\alpha-2}\right)$, which are of lower order than $\mathcal{O}\left(M\omega\right)$ (since $\alpha-2<1$). However, we will show that such corrections do not actually modify the final result for any $\alpha > 1$, and therefore Eq.~\eqref{eqKvalue} remains valid at order $\mathcal{O}\left(M\omega\right)$ even in this case.
\end{itemize}
In order to prove the second item, we need to consider the following equation
\begin{align}\label{eqordalpha}
  & y''(z)+ \left(1-\frac{l (l+1)}{z^2}\right) y(z)+ \notag\\&\quad \quad\quad \quad \quad+ A\frac{\left(M \omega\right)^{\alpha-2}}{z^{\alpha}}y(z)+\mathcal{O}\left(M\omega\right)=0\,. 
\end{align}
We observe that if $y(z,\omega)$ is a solution to the above equation, then necessarily $y(z e^{2\pi i}, \omega e^{2\pi i})$ is also a solution, since the equation remains invariant under the simultaneous transformation $z \to z e^{2\pi i}$ and $\omega \to \omega e^{2\pi i}$. 
Since we are dealing with a second-order differential equation, two possibilities arise:
\begin{itemize}
    \item $y(z e^{2\pi i}, \omega e^{2\pi i})$ is proportional to the original solution $y(z,\omega)$, up to a global multiplicative factor.
    \item $y(z e^{2\pi i}, \omega e^{2\pi i})$ picks up a contribution from the second linearly independent solution. This is exactly the case of Eq.~\eqref{eqKvalue}.
\end{itemize}
We are clearly interested in the second possibility. If the second case holds, then at asymptotic infinity, the ingoing and outgoing waves should mix. For $2<\alpha<3$, Eq.~\eqref{eqordalpha} admits the asymptotic solution
\begin{align}
    &y(z) \sim e^{\pm i z} \left(1 \mp \frac{i (l^2+l)}{2z} \right.\notag\\&\quad\quad \quad \quad \left. \pm \frac{i A (M\omega)^{\alpha-2}}{2 (\alpha -1) z^{\alpha-1}} + \mathcal{O}\left(\frac{1}{z^2}\right)\right)\,,
\end{align}
which does not exhibit any mixing between the ingoing and outgoing solutions under the simultaneous transformation $z \to z e^{2\pi i}$ and $\omega \to \omega e^{2\pi i}$. This is a sufficient condition to exclude mixing at all values of $z$. Indeed, if any mixing were present, it would also manifest at $z \to \infty$. As a consequence, this type of correction to the potential does not induce any discontinuity across the branch cut. Therefore, Eq.~\eqref{kvalue} remains valid, with mixing appearing only at order $\mathcal{O}\left(M\omega\right)$. 

It is worth noting that even when the potential is modified by terms of the form $A \log(r)/r^\alpha$, our argument still applies. One can show that such a term introduces corrections proportional to $A \omega^{\alpha-2} \log(z/\omega)/z^{\alpha-1}$ in the asymptotic solution, which remain invariant under the simultaneous transformation $z \to z e^{2\pi i}$ and $\omega \to \omega e^{2\pi i}$.

\subsection{Branch cut of $\psi_{r_-}$}\label{branchcutapp}
Consider Eqs.~\eqref{psirminus} and \eqref{branchcutpsiplus}. Clearly,
\begin{align}
    &\psi_{r_-}(r_*, \omega e^{2\pi i}) = A_{\rm out}(\omega) \left(\psi_{\infty_+}(r_*, \omega) - K(\omega)\, \psi_{\infty_-}(r_*, \omega)\right) \notag \\
    &\quad + A_{\rm in}(\omega) \left(\psi_{\infty_-}(r_*, \omega) - K(-\omega)\, \psi_{\infty_+}(r_*, \omega)\right)\,.
\end{align}
However, due to the way \( \psi_{r_-} \) is constructed (see Eq.~\eqref{psirminus_F}), it does not exhibit any branch cut, since it involves only the Coulomb wave function \( F_l(\eta, \rho) \), which is monodromic (see Appendix~\ref{connectionformulas}\footnote{{The inclusion of \( \mathcal{O}(M\omega) \) corrections, as carried out for \( \psi_{\infty_+} \), can be similarly applied to \( \psi_{r_-} \). In particular, the equation still admits a solution expressed as a series of Coulomb wave functions. However, the solution for \( \psi_{r_-} \) involves only the Coulomb function \( F_l(\eta, \rho) \), which is monodromic. Therefore, no branch cut arises in this case.
}}). It follows that the ingoing and outgoing wave amplitudes, \( A_{\rm in}(\omega) \) and \( A_{\rm out}(\omega) \), are polydromic functions of \( M\omega \). In particular, from Eq.~\eqref{psirminus} one obtains:
\begin{align}
    A_{\rm out}(\omega e^{2\pi i}) &= A_{\rm out}(\omega) - K^*(\omega)\, A_{\rm in}(\omega)\,, \\
    A_{\rm in}(\omega e^{2\pi i}) &= A_{\rm in}(\omega) + K(\omega)\, A_{\rm out}(\omega)\,.
\end{align}
Nonetheless, when computing the Green function in the low-frequency limit \( \omega \to 0 \), these corrections become subleading and vanish in the expression of Eq.~\eqref{Greenomegalow}. However, the above considerations will become relevant when computing the Fourier transform of the reflectivity (see Sec.~\ref{Refl_sec}).

\section{Numerical methods}\label{Numerics}
In order to verify our findings numerically, we solve Eq.~\eqref{waveeq} for an initially localized Gaussian wave packet. In particular, we adopt the following initial conditions in the tortoise coordinate:
\begin{equation}
    \frac{\partial}{\partial t} \Psi(r_*,t) = \begin{cases}
        e^{-\frac{(r_*-r_g)^2}{\delta^2}} \quad r_h < r_* < r_h + 5\delta\,, \\
        0 \quad r_* > r_h + 5\delta\,,
    \end{cases}
\end{equation}
with $\Psi(r_*,t) = 0$ initially. 

We set $r_g = 10M$ and $\delta = 5M$ for the cases discussed in Sec.~\ref{bhcase}. For the BH surrounded by matter profiles (Sec.~\ref{dmcase}), we localize the Gaussian packet at the same radial coordinate value across the different cases ($r \sim 14M$ for the cases shown), again using $\delta = 5M$. These choices produce tails scaling as $\propto t^{-2l-3}$. In all cases shown, the signal is extracted at $r_* \ll t-t'$ (for instance, $r_* \sim 50M$), and since for a Gaussian packet it also holds that $r_*' \ll t-t'$, the setup exactly matches the Price limit. We verified, by performing a resolution comparison, that the minimum strain amplitude not significantly affected by numerical noise in our simulations is approximately $10^{-12}$. However, in all the cases analyzed in this work, the strain remains larger than $10^{-9}$, ensuring that our results are not impacted by numerical artifacts.

When solving Eq.~\eqref{waveeq}, not only initial conditions but also boundary conditions must be specified.

In the BH case, we impose purely ingoing boundary conditions at the horizon, corresponding to
\begin{equation}
    \partial_{r_*}\Psi\big|_{r_h} = \partial_t\Psi\big|_{r_h}\,.
\end{equation}

On the other hand, when dealing with  horizonless objects, the boundary conditions must be properly modified. If the object is sufficiently compact, its surface can be modeled as $r_0 = r_h(1+\epsilon)$. For sufficiently small $\epsilon$, Eq.~\eqref{waveeq} admits plane wave solutions near the object's surface. Defining $u = r_* - t$ and $v = r_* + t$, a general solution can be written as
\begin{equation}
    \Psi(r_* \to r_{*,0}) = F(u) + G(v)\,,
\end{equation}
with
\begin{equation}
    \partial_{r_*} \Psi(r_* \to r_{*,0}) = F'(u) + G'(v)\,,
\end{equation}
and
\begin{equation}
    \partial_t \Psi(r_* \to r_{*,0}) = -F'(u) + G'(v)\,.
\end{equation}

By imposing
\begin{equation}
    \partial_{r_*} \Psi(r_* \to r_{*,0}) = \sigma\, \partial_t \Psi(r_* \to r_{*,0})\,,
\end{equation}
one finds that:
\begin{itemize}
    \item $\sigma = 1$ corresponds to purely ingoing waves at the surface (BH case).
    \item $\sigma = -1$ corresponds to purely outgoing waves at the surface.
    \item $\sigma = 0$ corresponds to a perfectly reflective surface at the radius.
\end{itemize}
In general, $0 < |\sigma| < 1$ describes a superposition of ingoing and outgoing waves at the surface. Although in this setup $\sigma$ is not directly related to the physical surface reflectivity, it is interesting to explore these boundary conditions, as they effectively model the boundary conditions expected for an exotic compact object.

In Fig.~\ref{fig:ECOtails}, we show the time-domain signal for several different values of $\sigma$.
\section{Perturbations of the Kerr metric}\label{KerrApp}
Kerr perturbations can be described by Eq.~\eqref{Kerr_waveeq}.  In order to factorize a purely ingoing wave at the horizon
\begin{equation}
 _{s}R_{lm}(r)=f(r)^{-2iM\tilde{\omega}} \,_{s}w_{lm}(r)\,,
\end{equation}
where $F(r)$ and $\tilde{\omega}$ are defined in Sec.~\ref{spinningcase}. At leading order in $M\omega$, Eq.~\eqref{Kerr_waveeq} reads 
\begin{equation}\label{Kerreq_worder0}
_{s}w_{lm}''(z) + \left(1 + \frac{2is}{z} - \frac{l(l+1)}{z^2} \right) \, _{s}w_{lm} + \mathcal{O}(M\omega) =0\,.
\end{equation}
It follows that two independent solutions to Eq.~\eqref{Kerr_waveeq} are given by 
\begin{align}
 & \Psi_1 = F(r)^{-2iM\tilde{\omega}}(r\omega)^{l+1}e^{ir\omega}\Phi(l+s+1,2l+2,-2ir\omega) \,,\notag\\ &  \Psi_2 = F(r)^{-2iM\tilde{\omega}}(r\omega)^{l+1}e^{ir\omega}U(l+s+1,2l+2,-2ir\omega)\,.
\end{align}
$\Phi(a,b,z)$ and $U(a,b,z)$ are Kummer functions. If one considers that 
\begin{equation}
    \lim_{z\to0}\Phi(a,b,z)=1\,
\end{equation}
it is easy to see that in the small frequency limit ($\Phi\omega\to0$) at the horizon ($r=r_+$), $\Psi_1$ satisfies the boundary condition of a purely ingoing wave, up to a normalization constant. Moreover, it holds that 
\begin{equation}
    \lim_{|z|\to \infty}U(a,b,z) \sim z^{-a}+\mathcal{O}(z^{-a-1})\,.
\end{equation}
Consequently, in the limit \( \omega \to 0 \) and \( r \to \infty \), the function \( \Psi_2 \) behaves as a purely outgoing wave at spatial infinity, up to a normalization constant. We now turn our attention to how these solutions transform under the analytic continuation \( \omega \to \omega e^{2\pi i} \). In particular, the function \( U(a,b,z) \) is known to be multi-valued, as discussed in Appendix~\ref{connectionformulas}. However, for the current values of \( a \) and \( b \), the branch cut does not manifest itself. Nonetheless, when Eq.~\eqref{Kerr_waveeq} is considered at order \( \mathcal{O}(M\omega) \), the discontinuity across the branch cut becomes relevant and must be properly taken into account.
By including the \( \mathcal{O}(M\omega) \) corrections, and introducing the redefinition \( _{s}R_{lm} = \psi / \sqrt{F(r)} \) along with the change of variable \( z = \omega r \), Eq.~\eqref{Kerr_waveeq} takes the form
\begin{align}
   &_{s}w_{lm}''(z) + \left(1 + \frac{2is}{z} - \frac{l(l+1)}{z^2} \right) \, _{s}w_{lm}  \notag+ {4 M \omega \over z}\psi (z)\\&\quad+M\omega\left( {k^{(2)}_{lm}(a)\over z^2}+{k^{(3)}_{lm}(a) \over z^3}\right)\psi (z)=\mathcal{O}\left((M\omega)^2\right)\,,
\end{align}
where
\begin{align}
    &k^{(2)}_{lm}(a)=\left(\frac{2 a m s^2 }{Ml (l+1)}-2 {a\over M} m +2 i  s\right) \notag\,,\\&k^{(2)}_{lm}(a)=\left(2 i {a\over M} m s-2 l(l+1)\right)\,.
\end{align}
As discussed in Sec.~\ref{branchcutapp} for the Schwarzschild case, at linear order in \( M\omega \), terms in the potential of the form \( 1/z^\alpha \) with \( \alpha > 1 \) do not contribute to the mixing between ingoing and outgoing wave solutions at infinity. One can show, by exploiting the properties of the confluent hypergeometric equation, that when only the modification of the \( 1/z \) term is taken into account, a pair of independent solutions is given by~\cite{Hod:1999ci}
\begin{equation}
  \psi_1(z) =(z)^{l+1}e^{iz}\Phi(l+s+1+2iM\omega,2l+2,-2iz) \,,
\end{equation}
and
\begin{equation}
    \psi_1(z) =(z)^{l+1}e^{iz}U(l+s+1+2iM\omega,2l+2,-2iz)\,.
\end{equation}
By implementing Eq.~\eqref{branchUM}, for our case we get
\begin{align}\label{eqBranchKerr}
    &\psi_2(z e^{2\pi i})=\psi_1(z) +\notag\\&\quad \quad(-1)^{l-s} { (l-s)!  \over (2l+1)!}4\pi M\omega\, \psi_2(z)\,.
\end{align}
The previous relation can be used in the computation of the Green function along the branch cut.
\bibliography{biblio}
\end{document}